\documentclass[journal]{IEEEtran}
\usepackage[T1]{fontenc}
\usepackage[latin9]{inputenc}
\usepackage{color}
\usepackage{float}
\usepackage{amsthm}
\usepackage[nosumlimits]{amsmath}
\usepackage{amssymb}
\usepackage{flushend}
\usepackage{multirow}
\usepackage{overpic}

\usepackage{hyperref}
\usepackage{breakurl}

\makeatletter

\floatstyle{ruled}
\newfloat{algorithm}{tbp}{loa}
\providecommand{\algorithmname}{Algorithm}
\floatname{algorithm}{\protect\algorithmname}

\theoremstyle{plain}

\theoremstyle{definition}

\theoremstyle{plain}

\theoremstyle{remark}
\newtheorem{rem}{\protect\remarkname}
\theoremstyle{remark}

\allowdisplaybreaks

\usepackage{mathtools}

\usepackage{amsfonts}
\usepackage{cite}
\usepackage{array}
\usepackage{algorithm}
\usepackage{algorithmic}
\usepackage{graphicx}
\usepackage{subfigure}

\DeclareMathOperator{\tr}{tr}

\DeclareMathOperator*{\maxi}{max}

\DeclareMathOperator*{\st}{s.t.}

\DeclarePairedDelimiter\ceil{\lceil}{\rceil}

\newcommand{\herm}{^{H}}
\newcommand{\trans}{^{T}}
\renewcommand{\Im}{\mathrm{Im}}
\renewcommand{\Re}{\mathrm{Re}}

\newcommand*{\rom}[1]{\expandafter\@slowromancap\romannumeral #1@}

\makeatother

\providecommand{\definitionname}{Definition}
\providecommand{\factname}{Fact}
\providecommand{\remarkname}{Remark}
\providecommand{\theoremname}{Theorem}
\providecommand{\lemmaname}{Lemma}

\mathchardef\mhyphen="2D

\begin{document}
\title{Energy-Efficient Beam Coordination Strategies with Rate Dependent Processing Power }

\author{Oskari Tervo,~\IEEEmembership{Student~Member,~IEEE}, Antti Tölli,~\IEEEmembership{Senior~Member,~IEEE}, Markku Juntti,~\IEEEmembership{Senior~Member,~IEEE}, and Le-Nam Tran,~\IEEEmembership{Member,~IEEE}
\thanks{This work was supported in part by Infotech Oulu Doctoral Program and the Academy of Finland under project Message and CSI Sharing for Cellular Interference Management with Backhaul Constraints (MESIC) belonging to the WiFIUS program with NSF. It has
also been co-funded by the Irish Government and the European Union under Irelands EU Structural and Investment Funds Programmes 2014-2020 through the SFI Research Centres Programme under Grant 13/RC/2077.}
\thanks{O. Tervo, A. Tölli and Markku Juntti are with Centre for Wireless Communications, University of Oulu,
Finland. Email: \{oskarite, antti.tolli, markku.juntti\}@ee.oulu.fi.}
\thanks{L.-N. Tran is with the Department of Electronic Engineering, Maynooth University, Ireland. Email: lenam.tran@nuim.ie.}}

\maketitle

\begin{abstract}
This paper proposes energy-efficient coordinated beamforming strategies for multi-cell multi-user multiple-input single-output system. We consider a practical power consumption model, where part of the consumed power depends on the base station or user specific data rates due to coding, decoding and backhaul. This is different from the existing approaches where the base station power consumption has been assumed to be a convex or linear function of the transmit powers. Two optimization criteria are considered, namely network energy efficiency maximization and weighted sum energy efficiency maximization. We develop successive convex approximation based algorithms to tackle these difficult nonconvex problems. We further propose decentralized implementations for the considered problems, in which base stations perform parallel and distributed computation based on local channel state information and limited backhaul information exchange. The decentralized approaches admit closed-form solutions and can be implemented without invoking a generic external convex solver. We also show an example of the pilot contamination effect on the energy efficiency using a heuristic pilot allocation strategy. The numerical results are provided to demonstrate that the rate dependent power consumption has a large impact on the system energy efficiency, and, thus, has to be taken into account when devising energy-efficient transmission strategies. The significant gains of the proposed algorithms over the conventional low-complexity beamforming algorithms are also illustrated.
\end{abstract}


\section{\label{sec:intro}Introduction}

Multi-antenna technology has been a core underlying component in modern wireless communication systems and will certainly remain its vital role in the development of the future 5G networks. Although the multi-antenna techniques have been shown to provide huge spectral efficiency gains, they cause a serious concern over the increased power consumption due to the number of associated radio frequency (RF) elements. In addition to the power directly used to transmit data, a significant amount of indirect power is consumed for other related operations, e.g., for running the base stations (BSs), baseband processing, RF processing, coding, decoding, and backhaul operations. In fact, in densely deployed networks, the data transmit power may be only a small part of the total power consumption according to 5G visions \cite{Osseiran-14}. As a result, energy efficiency (EE) has become an important design criterion for future networks \cite{Isheden:Framework:EE:2012,Ng:EE:OFDMA:2012,Venturino-15,Jiang-11,Tervo-15,Nguyen-15}. On the contrary to the conventional design criteria of either maximizing the sum rate or minimizing the required transmit power, energy efficiency optimization is to maximize the ratio between sum rate and total power consumption.

The main challenge in multiuser wireless communication systems design is due to multiuser interference caused by the use the same transmission resources. To this end, several interference coordination techniques have been proposed in the last decades. The idea of these methods is to mitigate the inter-cell interference by allowing cooperation between nearby cells. Among those, a powerful method adopted in current LTE-A systems is called \emph{coordinated beamforming}, where base stations can jointly design their beamforming vectors without sharing their data \cite{Irmer-11}. Another widely studied technique is joint transmission, where the data can be coherently transmitted from multiple base stations \cite{Ng-12}. However, since joint transmission requires a tight synchronization between the base stations, the focus of this paper is on the coordinated beamforming. It has been widely studied, e.g., for sum rate maximization \cite{Shi:WMMSE:2011,Komulainen-13,Tran-12}, and transmit power minimization \cite{Dahrouj:CoordinatedBF:2010,Tolli-11,Pennanen-11}. The energy-efficient coordinated beamforming strategy is highly dependent on the processing power resulting from the circuits and coordination operations of the base stations.


\subsubsection*{Related Work} The energy efficiency maximization (EEmax) problems belong to the class of fractional programs, which have been widely studied for both single-cell (SC) \cite{Tervo-15,Jiang-11,Wu-16} and multi-cell (MC) system models \cite{Nguyen-15,He-13,Li-14CoMP,Ng-12}. The problem of maximizing the minimum energy efficiency among base stations in a multi-cell multiuser multiple-input single-output (MISO) system was studied in \cite{Nguyen-15}. Energy-efficient joint transmission for multi-cell OFDMA systems with limited backhaul capacity and single-antenna base stations was considered in \cite{Ng-12}, where the power and subcarrier allocations were jointly optimized by assuming zero-forcing beamformers over all the base stations. Coordinated beamforming for network EEmax in multi-cell multi-antenna systems was studied in \cite{He-13,Li-14CoMP} where the latter incorporated the data rate constraints of users to the optimization problem. The weighted sum energy efficiency (WsumEE) was suggested as a performance measure in \cite{He-14,TervoWsumEE-15,Wu-16}, to satisfy the heterogeneous energy efficiency requirements of different cells or users. However, all these works only considered the circuit power as a constant and failed to recognize the fact that this sort of power consumption heavily depends on the data rate. More specifically, the circuit power is
an increasing function of the transmission rate, since a higher data rate requires a larger codebook which
incurs higher power for encoding and decoding on baseband
circuit boards \cite{Zhang-10,Ranpara-99,Bjornson-15,Wang-13,Wang-15,Arafa-15,Tutuncuoglu-12,Mahdavi-Doost-13,Grover-11,Rubio-13,Rost-10,Ganesan-11}.

The rate dependent power has been assumed to be either linear or non-linear convex increasing function of the data rate \cite{Zhang-10,Ranpara-99,Bjornson-15,Wang-13,Wang-15,Arafa-15,Tutuncuoglu-12,Mahdavi-Doost-13,Grover-11,Rubio-13,Rost-10,Ganesan-11}. The linear case with uniform user rates in a single-cell system was investigated in \cite{Bjornson-15}, where zero-forcing beamforming with massive MIMO setup was shown to achieve maximal energy efficiency. However, in a multi-cell network where the inter-cell interference experienced by each user becomes significant, zero-forcing method is highly suboptimal because the degrees of freedom are used up for nulling both intra- and inter-cell interference. A general convex power consumption model in point-to-point MIMO orthogonal frequency-division multiplexing systems was considered in \cite{Wang-13,Wang-15}, and in single-antenna energy-harvesting systems in \cite{Arafa-15,Mahdavi-Doost-13,Rubio-13}. The fact that the rate dependent power consumption can increase more than linearly with the data rate is shown, e.g., in \cite{Rost-10}, where the power consumption of the decoding process is investigated by using convolutional codes. Specifically, these codes can be decoded using a trellis representation of the encoder's state space, and the authors in \cite{Rost-10} show that the decoding complexity for each time symbol increases exponentially with the information rate. This complexity result is caused by the fact that the state space and the number of possible state transitions per channel access in the decoder-trellis expands exponentially with the product of constraint length and the information rate.

\subsubsection*{Contributions}
We study energy-efficient coordinated beamforming in multi-cell multi-user MISO systems with a general non-linearly increasing convex rate dependent power consumption model as in \cite{Wang-13,Wang-15}. This is different from the related research, which adopts either a simple (i.e., rate-independent) power consumption model \cite{He-14,TervoWsumEE-15,He-13} or a simplified beamforming technique (i.e., single-cell zero-forcing \cite{Bjornson-15} or a point-to-point system \cite{Wang-13,Wang-15}). Because the power consumption model depends on a specific implementation (e.g., the decoder), we use a general model to emphasize that the proposed algorithms can be applied to any power consumption model which is a convex increasing function of the data rate. Two different optimization criteria are considered. The first one is the network energy efficiency maximization (NetEEmax) which yields the maximum achievable energy efficiency of the network. The second one is the weighted sum energy efficiency maximization (WsumEEmax), which maximizes the sum of energy efficiencies of the cells. The latter is particularly relevant in heterogeneous networks, since it can balance or adjust the energy efficiencies and data rates between the cells and can be implemented in a decentralized manner \cite{TervoWsumEE-15}. The framework for the proposed solutions is based on successive convex approximation (SCA) principle \cite{Beck:SCA:2010}, which has been successfully applied in various wireless communications problems \cite{Tervo-15,Venkatraman-16,Kaleva-16}. The SCA is a local optimization method, where the main idea is to approximate the nonconvex part of the problem by proper convex bounds. In this way, the SCA method results in a sequence of convex subproblems that guarantee the feasibility of the iterates and monotonicity of the objective function. For the NetEEmax problem, we propose equivalent transformations to arrive at an iterative algorithm where a concave-convex fractional program is solved in each iteration using the Charnes-Cooper transformation \cite{Schaible-76}. The WsumEEmax problem is reformulated to derive an iterative method where a convex program is solved in each iteration. We also propose decentralized implementations in which base stations perform parallel and distributed computation based on local channel state information and limited backhaul information exchange. More specifically, we first reformulate the original problems by using the relation between signal-to-interference-plus-noise ratio (SINR) and mean squared error (MSE) when minimum MSE (MMSE) receiver is used. Then, we derive the optimality conditions for the approximated problem and propose an algorithm which combines the SCA and alternating optimization methods, and admits closed-form solutions.

To illustrate the pilot contamination effect on the energy efficiency, we propose a simple heuristic energy-efficient pilot allocation strategy. The idea of the proposed low-complexity method is to use (known) path gain information to calculate group-specific energy efficiency metrics and greedily allocate the pilot resources to the groups.

The numerical results illustrate that the proposed algorithms can provide significant energy efficiency gains (up to 60 \% in the considered setting) over the methods where the rate dependent power is not taken into account, showing the importance of including it in the optimization framework. We numerically compare the proposed algorithms with various conventional transmission schemes, and show that our proposed algorithms outperform all the existing coordinated beamforming designs in terms of energy efficiency.

Parts of this paper have been published in our previous conference publication \cite{Tervo-16}. The following additional contributions can be found in the present paper. We propose decentralized closed-form designs for the problems and also provide additional implementation aspects for the centralized solutions. We have also extended the system model to take into account the pilot overhead and pilot contamination, and consider more specific power consumption model. Furthermore, we propose a heuristic pilot allocation algorithm to address the problem of pilot contamination. We also provide an alternative iterative second-order cone program (SOCP) approximation algorithm to solve the WsumEEmax problem. Finally, we provide more extensive simulations to illustrate the performance of the methods.

\subsubsection*{Organization and Notation}
The rest of the paper is organized as follows. Section \ref{sec:ProblemFormulation} presents the system model, power consumption model and the considered optimization problems. The centralized algorithms are provided in Section \ref{sec:CentralizedMethods}, followed by the
decentralized methods in Section \ref{sec:DecentralizedMethods}. The computational complexity is discussed in Section \ref{ComplexityComp}. Pilot allocation strategy is proposed in Section \ref{sec:PilotAllocation} while numerical evaluation and conclusions are presented in Section \ref{sec:NumericalResults} and \ref{sec:Conclusions}, respectively.

The following notations are used in this paper. $|x|$ denotes the cardinality of $x$ if $x$ is a set, and absolute value of $x$, otherwise. $||\mathbf{x}||_2$ is a $\ell_{2}$ or Euclidean norm of $\mathbf{x}$ and boldcase letters without any superscripts or subscripts denote the set of variables. Otherwise, boldcase letters are vectors. $\mathbf{x}\trans, \mathbf{x}\herm, \Re(\mathbf{x})$ and $\Im(\mathbf{x})$ mean transpose, Hermitian transpose, real part and imaginary part of $\mathbf{x}$, respectively. $\tr(\mathbf{X})$ means trace of $\mathbf{X}$.

\section{\label{PF} System Model and Problem Formulation}
\label{sec:ProblemFormulation}
\subsection{System Model}
We consider the downlink (DL) of a multi-cell multiple-input single-output (MISO) system with $B$ cells. Each base station $b \in \mathcal{B} = \{1,\ldots,B\}$ equipped with $N_b$ antennas transmits data to a group of $K_b$ single-antenna users in its cell, represented by the set $\mathcal{K}_b$.
Each user $k \in \mathcal{K}\triangleq\cup_{b\in\mathcal{B}}\mathcal{K}_b$ in the network is served only by a single BS which is denoted by $b_k \in \mathcal{B}$, i.e., $\mathcal{K}_b \cap \mathcal{K}_{b'} = \emptyset \ \forall b \neq b'$.

In the downlink, the data symbol $s_{k}$ intended for
user $k$ is multiplied with the beamforming vector
$\mathbf{w}_{k}\in\mathbb{C}^{N_b\times1}$ before being transmitted.
Accordingly, the received signal at user $k$ is given by
\begin{eqnarray}
y_{k} & = & \mathbf{h}_{b_k,k}\herm\mathbf{w}_{k}s_{k}+\sum_{j\in\mathcal{K} \setminus \{k\}}\mathbf{h}_{b_j,k}\herm\mathbf{w}_{j}s_{j} +n_{k}\label{eq:signal:model}
\end{eqnarray}
where $\mathbf{h}_{b,k}\in\mathbb{C}^{N_b\times 1}$ is the channel vector from BS $b$ to user $k$, and $n_{k}$ is the background noise with distribution $\mathcal{CN}(0,\sigma^2)$. The data streams are assumed to be independent and have zero mean and unit power. As a result, we can write the signal-to-interference-plus-noise ratio of user $k$ as
\begin{equation}
\Gamma_{k}(\mathbf{w})\triangleq\frac{|\mathbf{h}_{b_k,k}\herm\mathbf{w}_{k}|^{2}}{N_0+\sum_{j\in\mathcal{K} \setminus \{k\}}|\mathbf{h}_{b_j,k}\herm\mathbf{w}_{j}|^2}
\end{equation}
where $\mathbf{w}\triangleq\{\mathbf{w}_{k}\}_{k\in \mathcal{K}}$ and $N_0=\sigma^2W$ by assuming that the system operates over bandwidth $W$ Hz.

We consider a block transmission, where the user channels stay constant within time-frequency coherence blocks of $U=W_\text{C}T_\text{C}$ channel uses/symbols. $W_\text{C}$ (in Hz) denotes the coherence bandwidth and $T_\text{C}$ (in second) is the coherence time. For the notational simplicity, we assume $W_C=W$ in this paper. We assume time-division duplex (TDD) protocol where all the cells are operating in the downlink mode. The protocol is matched to the coherence block so that each block consists of downlink demodulation pilots, uplink (UL) pilots, and downlink data transmission.
The BSs can estimate the downlink channels from the uplink pilots by exploiting the reciprocity. On the other hand, the users can estimate the effective channels (i.e., the combination of beamformer and channel) from the downlink demodulation pilots to decode the data. Let $\tau^{\text{ul}} \geq \underset{b}\max \ K_b$ and $\tau^{\text{dl}} \geq \underset{b}\max \ K_b$ be the total number of pilot resources in the network in uplink and downlink, respectively. This means that the total number of pilots in the coherence block is $\tau^{\text{ul}}+\tau^{\text{dl}}$. It is assumed that all the users are active across the transmission bandwidth, i.e., there is no frequency-domain scheduling. Thus, the rate expression of user $k$ for a given (known) channel realization is given
by
\begin{equation}\label{rateexp}
R_{k}(\mathbf{w})= (1-\tfrac{\tau^{\text{ul}}+\tau^{\text{dl}}}{U})W\log(1+\Gamma_{k}(\mathbf{w}))
\end{equation}
where $(1-\tfrac{\tau^{\text{ul}}+\tau^{\text{dl}}}{U})$ accounts for the pilot overhead in UL and DL.\footnote{The methods are easily extended to a frequency-selective case where the rate would be defined as $R_{k}(\mathbf{w})= (1-\tfrac{\tau^{\text{ul}}+\tau^{\text{dl}}}{U})\sum_{m\in\mathcal{M}}W_C\log(1+\Gamma_{k,m}(\mathbf{w}_m))$, where $\mathcal{M}$ is the set of coherence bands and the SINR expressions are per coherence band.}As is well-known, the capacity expressions in information theory are based on the length of the codewords approaching infinity. Although this can never be exactly realized in practice, the state of the art powerful codes such as turbo, low density parity check (LDPC) and polar codes \cite{Balatsoukas-17,Dahlman-16} can approximately achieve the bounds with practical codeword block sizes (of less than 10000 bits) at low to moderate SNR levels with the accuracy sufficient for practical purposes.\footnote{For example, with 100 Mbit/s high rate mobile broadband, this would require the channel to remain static for a block size of 100 $\mu s$ in time domain, which is not a hard assumption.} In that sense, \eqref{rateexp} is a practically relevant upper bound for the achievable rate. In other words, we focus on block transmission wherein the block is large enough in terms of data bits such that the information theoretic rate expression is a relevant upper bound for the achievable rate within a block of static channel state over coherence time; this rate value is deterministic given the known channel realization. Simultaneously, the block needs to be short enough in physical time units such that the channel remains constant over the block. Note that the optimization is performed for a fixed channel. When the final performance is evaluated via computer simulations, the rate bound in \eqref{rateexp} can be averaged out over the fading channel states.


\subsubsection*{Pilot contamination}
The pilot reuse results in the effect of pilot contamination. For simplicity, we focus on uplink pilot contamination and assume that the users have perfect effective channel information, i.e, we set $\tau^{\text{dl}}=|\mathcal{K}|$ throughout the paper. However, in practice the pilot reuse is meaningful for the downlink demodulation pilots as well. If $\tau^{\text{ul}} = |\mathcal{K}|$, the BSs have perfect channel information towards all the users, since it allows to allocate orthogonal pilot resources to each user. On the other hand, if $\tau^{\text{ul}} < |\mathcal{K}|$, then some of the pilot resources need to be reused, which leads to the effect of pilot contamination. Assuming that the channels are perfectly estimated by each BS from the received uplink pilots, which are contaminated by the pilot signals of the users using the same pilot resources, the observed channel at BS $b$ becomes
\begin{equation}
\tilde{\mathbf{h}}_{b,k} = \hat{\mathbf{h}}_{b,k} + \sum_{j\in\mathcal{K}^i}\hat{\mathbf{h}}_{b,j}
\end{equation}
where $\hat{\mathbf{h}}_{b,k}\in\mathbb{C}^{N_b\times 1}$ is the (perfect) channel vector from BS $b$ to user $k$, and $\mathcal{K}^i$ is the set of users using pilot resource $i$. The above expression uses a standard assumption of high-SNR pilots so that the noise term vanishes. As a result, the BS performs the beamformer optimization based on $\tilde{\mathbf{h}}_{b,k}$, instead of $\hat{\mathbf{h}}_{b,k}$.
\subsection{Power Consumption Model}
We combine the power consumption models from \cite{Wang-15,Wang-13,Bjornson-15} and extend the general rate dependent power consumption model to account for the multi-cell multi-user transmission. As a result, the total power consumption of BS $b$ writes as
\begin{equation}
P_{\text{tot},b}=\dfrac{1}{\eta}\sum_{k\in\mathcal{K}_b}||\mathbf{w}_k||_2^2+P_{\text{CP},b} + P_{\text{RD}}\delta(r_b)\label{eq:powermodel}
\end{equation}
where the first term is the data transmit power in the downlink, $\eta\in[0,1]$ is the power amplifier
efficiency at the BS, $P_{\text{CP},b}$ is the total rate independent circuit power consumption of cell $b$, $P_{\text{RD}}\geq 0$ is a constant accounting for the coding, decoding and backhaul power consumption, and $\delta(r_b)$ is a differentiable, strictly increasing and convex function of the total sum rate $r_b$ of BS $b$, satisfying $\delta(0)=0$. The rate dependent power consumption of BS $b$ could also be modeled as $\sum_{k\in\mathcal{K}_b}P_{\text{RD}}\delta(r_k)$, where $\delta(r_k)$ is a function of individual user rate $r_k$. In Appendix \ref{app:altPC}, we show how to modify the algorithms proposed in the subsequent sections to deal with this alternative model. $P_{\text{CP},b}$ in \eqref{eq:powermodel} is decomposed as
\begin{equation}
P_{\text{CP},b}=P_{\text{FIX}}+P_{\text{TC},b}+P_{\text{CE}}+P_{\text{LP},b}\label{eq:CP}
\end{equation}
where $P_{\text{FIX}}$ is a fixed power consumption required for site-cooling, control signaling, and the load-independent power of backhaul infrastructure and baseband processors, $P_{\text{TC},b}$ is the power consumption of the transceiver chains, $P_\text{CE}$ is the power consumed in channel estimation for DL and UL, and $P_{\text{LP},b}$ is the power used for linear processing at the BS side. $P_{\text{TC},b}$ can be further decomposed as
\begin{equation}
P_{\text{TC},b}=N_bP_{\text{BS}}+P_{\text{SYN}}+K_bP_{\text{UE}}\label{eq:TC}
\end{equation}
where $P_{\text{BS}}$ is the power per RF chain at each antenna, $P_{\text{SYN}}$ is the power consumed by local oscillator and $P_{\text{UE}}$ is the fixed circuit power of each user. To model the power consumed for computation of the beamformers and linear processing, we exploit the model used in \cite{Bjornson-15}. In this regard, $P_{\text{LP},b}$ can be expressed as
\begin{equation}\label{eq:PLP}
P_{\text{LP},b} = W(1-\frac{(\tau^{\text{ul}}+\tau^\text{dl})}{U})\frac{2N_b|\mathcal{K}_b|}{L_{\text{BS}}} + P_\text{LP,c},
\end{equation}
where $L_{\text{BS}}$ [flops/W or flop/J] is the computational efficiency of a BS. The first term in \eqref{eq:PLP} is the power consumed for the linear multiplication of the beamformers and the data symbols over the whole transmission bandwidth, while the second term $P_\text{LP,c}$ is the power consumed for the computation of the beamformers. The power consumption in relation to beamformer optimization depends on the specific algorithm. Since all the proposed methods are iterative, we can write
\begin{equation}
P_\text{LP,c} = QP_\text{LP,iter},
\end{equation}
where $Q$ is the required number of iterations to calculate the beamformers. The value of $P_\text{LP,iter}$ depends on the complexity of each iteration and is discussed in Section \ref{ComplexityComp}.\footnote{The fixed power values in general can be different for different BSs especially in heterogeneous networks, but for simplicity they are assumed to be equal for all the cells.}
\subsection{Problem Formulation }
Let us define $\tilde{R}_b(\mathbf{w})\triangleq \alpha\sum_{k\in\mathcal{K}_b}\log(1+\Gamma_{k}(\mathbf{w}))$ to be a function denoting the total sum rate of BS $b$, where $\alpha\triangleq (1-\tfrac{\tau^{\text{ul}}+\tau^{\text{dl}}}{U})W$. The first problem considered in this paper is called network energy efficiency maximization stated as
\begin{subequations} \label{EEmax}
\begin{eqnarray}{} \underset{\mathbf{w}}{\maxi} &  & \frac{\sum_{b\in\mathcal{B}}\tilde{R}_b(\mathbf{w})}{g(\mathbf{w})+P_{\text{RD}}\sum_{b\in\mathcal{B}}\delta(\tilde{R}_b(\mathbf{w}))}  \label{eq:EE:obj} \\ \st
&    & \sum_{k \in \mathcal{K}_b}||\mathbf{w}_{k}||_{2}^{2}\leq P_b,\;\forall b \in \mathcal{B} \label{eq:EE:PC}
\end{eqnarray}
\end{subequations}
where $g(\mathbf{w}) \triangleq {\frac{1}{\eta}}\sum_{k\in\mathcal{K}}||\mathbf{w}_{k}||_{2}^{2} + \sum_{b\in\mathcal{B}}P_{\text{CP},b}$ includes the power consumption which does not depend on the rate function. Conventionally, the denominator of the objective function has been either linear or convex function of the power values \cite{Wang-13,Wang-15,Bjornson-15}. However, this assumption no longer holds for the problem in \eqref{EEmax}.

The second problem of interest is the one of weighted sum energy efficiency maximization with BS-specific power constraints written as
\begin{subequations}\label{WsumEEmax}
\begin{eqnarray}{} \underset{\mathbf{w}}{\maxi} &  & \sum_{b\in\mathcal{B}}\omega_b\frac{\tilde{R}_b(\mathbf{w})}{g_b(\tilde{\mathbf{w}}_b)+P_{\text{RD}}\delta(\tilde{R}_b(\mathbf{w}))}  \label{eq:WsumEE:obj} \\ \st
&    & \sum_{k \in \mathcal{K}_b}||\mathbf{w}_{k}||_{2}^{2}\leq P_b,\;\forall b \in \mathcal{B} \label{eq:WsumEE:PC}
\end{eqnarray}
\end{subequations}
where $\tilde{\mathbf{w}}_b\triangleq\{\mathbf{w}_k\}_{k\in\mathcal{K}_b}$, $g_b(\tilde{\mathbf{w}}_b) \triangleq {\frac{1}{\eta}}\sum_{k\in\mathcal{K}_b}||\mathbf{w}_{k}||_{2}^{2} + P_{\text{CP},b}$ refers to the power consumption which is independent of the rate function $\tilde{R}_b(\mathbf{w})$, and $\omega_b$ is the energy efficiency priority weighting factor for BS $b$. Despite the apparent similarity, the WsumEEmax problem is somewhat more difficult to tackle compared to \eqref{EEmax},  simply because the objective \eqref{eq:WsumEE:obj} is a sum of fractional functions, which is not quasiconcave even if the numerators and denominators are linear. The energy efficiency metrics \eqref{eq:EE:obj} and \eqref{eq:WsumEE:obj} above can be seen as achievable instantaneous EE values optimized per channel realization.
\section{Proposed Centralized Solutions}\label{sec:CentralizedMethods}

\subsection{Network Energy Efficiency Maximization}
We remark that the problem in \eqref{EEmax} is not a concave fractional program for which efficient methods are known \cite{Dinkelbach-67}, \cite{Schaible-76}. The obvious reason is that both the numerator and the denominator in \eqref{eq:EE:obj} are nonconvex. To find a more tractable reformulation, we introduce the following equivalent transformation of \eqref{EEmax}
\begin{subequations}
\label{eq:EEmax:reform1}
\begin{eqnarray}
\underset{\mathbf{w},\mathbf{r}}{\maxi} &  &  \frac{\sum_{b\in\mathcal{B}}r_b}{g(\mathbf{w}) + P_{\text{RD}}\sum_{b\in\mathcal{B}}\delta(r_b)}\\
\st &  &  r_b \leq \alpha\sum_{k\in\mathcal{K}_b}\log(1+\Gamma_k(\mathbf{w})), \forall b \in \mathcal{B}\label{eq:EEmax:reform1:rate}\\
&  &  \sum_{k \in \mathcal{K}_b}||\mathbf{w}_{k}||_{2}^{2}\leq P_b,\;\forall b \in \mathcal{B} \label{eq:EE:PC}
\end{eqnarray}
\end{subequations}
where \eqref{EEmax} and \eqref{eq:EEmax:reform1} are equivalent because the constraints in \eqref{eq:EEmax:reform1:rate} are active at optimality, and $\mathbf{r}\triangleq\{r_b\}_{b\in\mathcal{B}}$ are new variables representing the sum rate of each base station $b$.\footnote{The proposed algorithms can be straightforwardly extended to include data rate constraints, i.e., $r_b \geq R_{b}^{\text{target}}$ or $r_k \geq R_{k}^{\text{target}}$.} At this point, we note that the objective function has become a linear-convex fractional function and the difficulty in solving \eqref{eq:EEmax:reform1} is due to the constraint \eqref{eq:EEmax:reform1:rate}. To this end, we introduce  new variables $\boldsymbol\gamma\triangleq\{\gamma_k\}_{k\in\mathcal{K}}$ to represent the SINR of each user $k$, and equivalently formulate \eqref{eq:EEmax:reform1} as
\begin{subequations}
\label{eq:EEmax:reform2}
\begin{align}
& \hspace{-12mm} \underset{\mathbf{w},\mathbf{r},\boldsymbol\gamma}{\maxi} \quad \frac{\sum_{b\in\mathcal{B}}r_b}{g(\mathbf{w}) + P_{\text{RD}}\sum_{b\in\mathcal{B}}\delta(r_b)}\\
\st \quad & r_b \leq \alpha\sum\nolimits_{k\in\mathcal{K}_b}\log(1+\gamma_{k}), \forall b \in \mathcal{B}\\
\quad & \gamma_k \leq \frac{|\mathbf{h}_{b_k,k}\herm\mathbf{w}_{k}|^{2}}{N_0+\sum_{j\in\mathcal{K} \setminus \{k\}}|\mathbf{h}_{b_j,k}\herm\mathbf{w}_{j}|^2}, \forall k \in \mathcal{K}\label{eq:EEmax:reform1:SINR}\\
\quad &  \sum_{k \in \mathcal{K}_b}||\mathbf{w}_{k}||_{2}^{2}\leq P_b,\;\forall b \in \mathcal{B} \label{eq:EE:PC}
\end{align}
\end{subequations}
where SINR constraints in \eqref{eq:EEmax:reform1:SINR} are still nonconvex.
By introducing a new variables $\boldsymbol\beta\triangleq\{\beta_k\}_{k\in\mathcal{K}}$ for total interference-plus-noise of user $k$ \cite{Venkatraman-16},\cite{Tran-12}, we can further rewrite \eqref{eq:EEmax:reform2} as
\begin{subequations}
\label{eq:EEmax:reform3}
\begin{align}
& \hspace{-12mm} \underset{\mathbf{w},\mathbf{r},\boldsymbol\gamma,\boldsymbol\beta}{\maxi} \quad \frac{\sum_{b\in\mathcal{B}}r_b}{g(\mathbf{w}) + P_{\text{RD}}\sum_{b\in\mathcal{B}}\delta(r_b)}\\
\st \quad  & \gamma_k \leq |\mathbf{h}_{b_k,k}\herm\mathbf{w}_{k}|^{2}/\beta_k, \forall k\in\mathcal{K}\label{eq:quad_over_lin}\\
\quad  & r_b \leq \alpha\sum\nolimits_{k\in\mathcal{K}_b}\log(1+\gamma_{k}), \forall b \in \mathcal{B}\\
\quad  & \beta_k \geq N_0+\sum_{j\in\mathcal{K} \setminus \{k\}}|\mathbf{h}_{b_j,k}\herm\mathbf{w}_{j}|^2, \forall k\in\mathcal{K}\label{eq:interference:constraint}\\
\quad  & \sum_{k \in \mathcal{K}_b}||\mathbf{w}_{k}||_{2}^{2}\leq P_b,\;\forall b \in \mathcal{B} \label{eq:EE:PC}
\end{align}
\end{subequations}
which lends itself to the application of successive convex approximation framework \cite{Beck:SCA:2010,Marks:IA:1977}. Specifically the right hand side of \eqref{eq:quad_over_lin} is called a quadratic-over-linear function which is jointly convex with respect to $\beta_k$ and $\mathbf{w}_k$. Thus, we can use the first-order lower approximation for the right side of \eqref{eq:quad_over_lin} as \cite{TervoWsumEE-15}
\begin{eqnarray}\label{eq:EEmax:quad_over_lin_app}
 |\mathbf{h}_{b_k,k}\herm\mathbf{w}_{k}|^2/\beta_{k} \geq 2\Re((\mathbf{w}_{k}^{(n)})\herm\mathbf{h}_{b_k,k}\mathbf{h}_{b_k,k}\herm\mathbf{w}_{k})/\beta_{k}^{(n)}\nonumber \\
 - (|\mathbf{h}_{b_k,k}\herm\mathbf{w}_{k}^{(n)}|/\beta_{k}^{(n)})^2\beta_{k} \triangleq \Psi_k^{(n)}(\mathbf{w}_{k},\beta_{k}).
\end{eqnarray}
According to the SCA principle we will replace the right side of \eqref{eq:quad_over_lin} by a convex lower bound. From \eqref{eq:EEmax:quad_over_lin_app}, the problem at iteration $n$ of the proposed SCA-based algorithm can be expressed as
\begin{subequations}
\label{eq:EEmax:reform4}
\begin{align}
& \hspace{-12mm} \underset{\mathbf{w},\boldsymbol\gamma,\boldsymbol\beta,\mathbf{r}}{\maxi} \quad \frac{\sum_{b\in\mathcal{B}}r_b}{g(\mathbf{w}) + P_{\text{RD}}\sum_{b\in\mathcal{B}}\delta(r_b)}\label{eq:EEmax:reform4obj}\\
\st \quad  & \gamma_k \leq \Psi_k^{(n)}(\mathbf{w}_k,\beta_k), \forall k\in\mathcal{K}\label{eq:CC_quad_over_lin_approx}\\
\quad  & r_b \leq \alpha\sum\nolimits_{k\in\mathcal{K}_b}\log(1+\gamma_{k}), \forall b \in \mathcal{B} \label{eq:rlinear}\\
\quad  & \beta_k \geq N_0+\sum_{j\in\mathcal{K}\setminus \{k\}}|\mathbf{h}_{b_j,k}\herm\mathbf{w}_{j}|^2, \forall k\in\mathcal{K}\\
\quad  & \sum_{k \in \mathcal{K}_b}||\mathbf{w}_{k}||_{2}^{2}\leq P_b,\;\forall b \in \mathcal{B} \label{eq:EE:PC}
\end{align}
\end{subequations}
which is a concave-convex fractional program. The common way of solving concave-convex fractional program is to use the Dinkelbach method \cite{Dinkelbach-67} which requires iterative processing. Here, we use a parameter-free approach based on the Charnes-Cooper transformation \cite{Schaible-76}. Specifically, the concave-convex fractional program can be transformed to an equivalent convex program with the transformations $\bar{\mathbf{w}}_k=\phi\mathbf{w}_k,\bar{\gamma}_k=\phi\gamma_k,\bar{\beta}_k=\phi\beta_k,\bar{r}_b=\phi r_b$ and $\phi=\tfrac{1}{\frac{1}{\eta}\sum_{k\in\mathcal{K}}||\mathbf{w}_{k}||_2^2 + \sum_{b\in\mathcal{B}}(P_{\text{RD}}\delta(r_b) + P_{\text{CP},b})}$. As a result, solving \eqref{eq:EEmax:reform4} boils down to solving the following convex program
\begin{subequations}
\label{eq:EEmax:reformFinal}
\begin{align}
& \hspace{-9mm} \underset{\bar{\mathbf{w}},\bar{\boldsymbol\gamma},\bar{\boldsymbol\beta},\bar{\mathbf{r}},\phi}{\maxi} \quad \sum_{b\in\mathcal{B}}\bar{r}_b\\
\st \quad  & \bar{\gamma}_k - \Psi_k^{(n)}(\bar{\mathbf{w}}_k,\bar{\beta}_k)) \leq 0, \forall k\in\mathcal{K}\label{eq:CC_quad_over_lin}\\
\quad  & \bar{r}_b - \alpha\sum\nolimits_{k\in\mathcal{K}_b}\phi\log(1+\tfrac{\bar{\gamma}_{k}}{\phi}) \leq 0, \forall b \in \mathcal{B}\label{eq:exponential}\\
\quad  & \phi N_0+\sum_{j\in\mathcal{K}\setminus \{k\}}\tfrac{|\mathbf{h}_{b_j,k}\herm\bar{\mathbf{w}}_{j}|^2}{\phi} - \bar{\beta}_k \leq 0, \forall k\in\mathcal{K}\label{eq:IntConstForApp}\\
\quad  & \frac{1}{\eta}\sum_{k\in\mathcal{K}}\tfrac{||\bar{\mathbf{w}}_{k}||_2^2}{\phi} + \sum_{b\in\mathcal{B}}(P_{\text{RD}}\phi\delta(\tfrac{\bar{r}_b}{\phi}) + \phi P_{\text{CP},b}) \leq 1\label{eq:EEmax:reformFinal:TotPower}\\
\quad  & \sum_{k\in\mathcal{K}_b}\tfrac{||\bar{\mathbf{w}}_k||_2^2}{\phi} \leq \phi P_b, \forall b \in \mathcal{B}.\label{eq:PowConstForApp}
\end{align}
\end{subequations}
The optimal solutions for problem \eqref{eq:EEmax:reform4} can be extracted as $\mathbf{w}_{k}^*=\tfrac{\bar{\mathbf{w}}_{k}^*}{\phi^*},\gamma_{k}^*=\tfrac{\bar{\gamma}_{k}^*}{\phi^*}, \beta_{k}^*=\tfrac{\bar{\beta}_{k}^*}{\phi^*}, r_b^*=\tfrac{\bar{r}_b^*}{\phi^*}$. In the proposed algorithm, we iteratively approximate \eqref{EEmax} by \eqref{eq:EEmax:reformFinal} until convergence. The proposed algorithm is presented in Algorithm \ref{algo:iterative}. The monotonic convergence of the objective function for Algorithm \ref{algo:iterative} is not difficult to see. Specifically, due to linear approximation \eqref{eq:EEmax:quad_over_lin_app}, the constraints \eqref{eq:CC_quad_over_lin} become loose after each update of $\bar{\mathbf{w}}^{(n)},\bar{\boldsymbol\beta}^{(n)}$. Furthermore, the updating rules ensure feasibility of the next iteration. These facts guarantee that $\sum_{b\in\mathcal{B}}\bar{r}_b^{(n+1)}\geq \sum_{b\in\mathcal{B}}\bar{r}_b^{(n)}$. Finally, the power constraint bounds each $\bar{r}_b$ from above. More detailed convergence analysis of the sequence of iterates and convergence point for the problem with similar structure can be found in \cite[Appendix A]{Venkatraman-16}. We remark that Alg. \ref{algo:iterative} applies to any convex function $\delta(\cdot)$.

\begin{rem}\label{remPowerModel}
The function $\phi\delta(\tfrac{\bar{r}_b}{\phi})$ in constraint \eqref{eq:EEmax:reformFinal:TotPower} is a perspective transformation of $\delta(\bar{r}_b)$ which is also convex due to the convexity of $\delta(\bar{r}_b)$ \cite{Boyd:ConvexOpt:2004}. In practice, it is computationally efficient to reformulate \eqref{eq:EEmax:reformFinal:TotPower} in a form which can be handled by dedicated powerful convex solvers. This is entirely possible depending on the type of $\delta(\bar{r}_b)$. Specifically, if $\delta(y)=y$ or $\delta(y)=y^2$, then \eqref{eq:EEmax:reformFinal:TotPower} immediately admits a second-order cone (SOC) representation. Let us now consider a general power model considered in \cite{Wang-15} where $\delta(y)=y^m$ with constant $m>1$ is always convex in the domain $y\geq 0$. For all practical purposes we can assume that $m$ is a rational number without loss of optimality. Thus, there exist $k$ and $c$ such that $k\geq1$ and $m=c/k$. By introducing a slack variable $x_b$, we can equivalently express \eqref{eq:EEmax:reformFinal:TotPower} as
\begin{subequations}
\label{eq:perspectives}
\begin{align}
& \frac{1}{\eta}\sum_{k\in\mathcal{K}}\tfrac{||\bar{\mathbf{w}}_{k}||_2^2}{\phi} + \sum_{b\in\mathcal{B}}(P_{\text{RD}}x_b + \phi P_{\text{CP},b}) \leq 1\label{eq:EEmax:reformFinal:TotPowerReform1}\\
\quad  & -\bar{r}_b \geq - \phi^{1-\tfrac{k}{c}}x_b^{\tfrac{k}{c}}, \forall b \in \mathcal{B}.\label{eq:EEmax:reformFinal:Geomean}
\end{align}
\end{subequations}
Note that the constraint in \eqref{eq:EEmax:reformFinal:Geomean} is an inequality involving rational powers \cite[Eq. (11)]{Alizadeh:SOCP:2001} and can be implemented as a series of SOC constraints as shown after \cite[Eq. (11)]{Alizadeh:SOCP:2001}.
\end{rem}

\begin{rem}\label{Lemma1}
If the power consumption has a linear dependence on the rate, i.e., $\delta(r_b) = r_b$ and $P_{\text{RD}}$ is the same for all the BSs, then $P_{\text{RD}}\delta(r_b)$ does not affect the optimal variables of \eqref{EEmax}. In this special case, \eqref{eq:EE:obj} is equal to $\min({g(\mathbf{w})}/\sum_{b\in\mathcal{B}}\tilde{R}_{b}(\mathbf{w}) + P_{\text{RD}})$. As can be seen, $P_{\text{RD}}$ becomes a constant in the objective function and could be ignored in the optimization process without loss of optimality (but not in the actual utility). This means that the optimal beamformers of \eqref{EEmax} and the problem considered in \cite{He-13} would be equal in this special case.
\end{rem}

\begin{algorithm}[t]
\begin{algorithmic}[1] \caption{Proposed SCA-based beamformer design for the network energy efficiency maximization problem.}
\label{algo:iterative} \renewcommand{\algorithmicrequire}{\textbf{Initialization:}}
\REQUIRE Set $n=0$, and generate initial points $(\bar{\mathbf{w}}^{(n)},\bar{\boldsymbol\beta}^{(n)})$.
\REPEAT
\STATE Solve \eqref{eq:EEmax:reformFinal} with $(\bar{\mathbf{w}}^{(n)},\bar{\boldsymbol\beta}^{(n)})$
and denote optimal values as $(\bar{\mathbf{w}}^*,\bar{\boldsymbol\beta}^*)$.
\STATE Update $(\bar{\mathbf{w}}^{(n)} = \bar{\mathbf{w}}^* ,\bar{\boldsymbol\beta}^{(n)} = \bar{\boldsymbol\beta}^*)$.\label{algo:update}
\STATE $n:=n+1$.
\UNTIL desired accuracy
\renewcommand{\algorithmicrequire}{\textbf{Output:}} \REQUIRE $\mathbf{w}_k^* = \tfrac{\bar{\mathbf{w}}_k^*}{\phi^*}, \forall k \in \mathcal{K}$
\end{algorithmic}
\end{algorithm}

\subsection{Weighted Sum Energy Efficiency Maximization}
The objective function of \eqref{WsumEEmax} is a sum-of-ratios objective function. A common approach to solve a sum-of-ratios maximization problem with concave-convex ratios is to transform it to a parameterized form with some fixed parameters, and then search the optimal parameters by solving a series of convex subproblems \cite{He-14,Wu-16}. Specifically, the general form of sum-of-fractional program $\max_{\mathbf{x}}\sum\frac{f_i(\mathbf{x})}{g_i(\mathbf{x})}$ can be solved as a series of subproblems $\max_{\mathbf{x}}\sum_i\alpha_i(f_i(\mathbf{x})-\beta_i g_i(\mathbf{x}))$, where $\alpha_i,\beta_i$ are some parameters. At each iteration, $\alpha_i$ and $\beta_i$ are fixed. After solving the parameterized program, $\alpha_i,\beta_i$ are updated according to a damped Newton method \cite{Jong-12} (see \cite{Jong-12} for details). However, in problem \eqref{WsumEEmax}, both the numerator $f_i(\mathbf{x})$ and the denominator $g_i(\mathbf{x})$ in each EE function are non-convex, meaning that such a parametric approach would result in a multi-level iterative algorithm due to the non-convexity of each parameterized subproblem. To avoid this drawback, we propose an approach with only one iteration loop.

Here we again apply the SCA framework to solve \eqref{WsumEEmax}. As the first step, we introduce new variables $\mathbf{t}\triangleq\{t_b\}_{b\in\mathcal{B}}$ to represent the energy efficiency of each cell and arrive at the following equivalent transformation of \eqref{WsumEEmax}
\begin{subequations}
\label{eq:WsumEEmax:reform1}
\begin{align}
& \hspace{-12mm} \underset{\mathbf{t},\mathbf{w}}{\maxi} \quad \sum_{b\in\mathcal{B}} \omega_bt_b \label{eq:obj1}\\
\st \quad  & t_b \leq \frac{\tilde{R}_b(\mathbf{w})}{g_b(\tilde{\mathbf{w}}_{b}) + P_{\text{RD}}\delta(\tilde{R}_b(\mathbf{w}))}, \forall b \in \mathcal{B}\label{eq:epi1}\\
\quad  & \sum_{k \in \mathcal{K}_b}||\mathbf{w}_{k}||_{2}^{2}\leq P_b,\;\forall b \in \mathcal{B} \label{eq:EE:PC}
\end{align}
\end{subequations} which is in fact an epigraph form of \eqref{WsumEEmax}.
Next, by introducing new user-specific SINR variables $\{\gamma_k\}_{k\in\mathcal{K}}$ and BS-specific rate variables $\{r_b\}_{b\in\mathcal{B}}$ as in network EEmax problem, we can further equivalently reformulate \eqref{eq:WsumEEmax:reform1} as
\begin{subequations}
\label{eq:WsumEEmax:reform3}
\begin{align}
&\underset{\mathbf{t},\mathbf{w},\mathbf{r},\boldsymbol\gamma}{\maxi} \quad   \sum_{b\in\mathcal{B}} \omega_bt_b\\
&\st \hspace{20pt}   t_b \leq \frac{r_{b}^2}{g_b(\tilde{\mathbf{w}}_{b}) + P_{\text{RD}}\delta(r_b)}, \forall b \in \mathcal{B}\label{eq:epi3}\\
& \hspace{50pt}  \gamma_{k} \leq \Gamma_{k}(\mathbf{w})
, \forall k \in \mathcal{K} \label{eq:SINRnonconvex}\\
&  \hspace{50pt}\alpha\sum_{k\in\mathcal{K}_b}\log(1+\gamma_{k}) \geq r_{b}^2, \forall b \in \mathcal{B}\label{eq:logalpha}\\
& \hspace{50pt} \sum_{k \in \mathcal{K}_b}||\mathbf{w}_{k}||_{2}^{2}\leq P_b,\;\forall b \in \mathcal{B}.
\end{align}
\end{subequations} We have used $r_b^2$ in \eqref{eq:logalpha} rather than $r_b$ as in \eqref{eq:rlinear} so that we can directly present \eqref{eq:epi3} as a difference of convex (DC) constraint. The equivalence between \eqref{eq:WsumEEmax:reform3} and \eqref{eq:WsumEEmax:reform1} is guaranteed, since all the constraints \eqref{eq:epi3}-\eqref{eq:logalpha} are active at optimality.
The constraints in \eqref{eq:SINRnonconvex} are equivalent to \eqref{eq:EEmax:reform1:SINR} and can be handled as shown in \eqref{eq:EEmax:reform3} and \eqref{eq:EEmax:quad_over_lin_app}. To find a tractable reformulation of nonconvex constraints \eqref{eq:epi3}, we can equivalently split it into the following two constraints
\begin{subequations}\label{eq:EEmax:replacedconst}
\begin{align}
&  t_b \leq \frac{r_b^2}{z_b}, \forall b \in \mathcal{B}\label{eq:Wsum:noncvxorig1}\\
&  z_b \geq g_b(\tilde{\mathbf{w}}_{b}) + P_{\text{RD}}\delta(r_b), \forall b \in \mathcal{B}\label{eq:Wsum:cvxorig2},
\end{align}
\end{subequations}
where we have introduced new variables $\mathbf{z}\triangleq\{z_b\}_{b\in\mathcal{B}}$.
Now, \eqref{eq:Wsum:cvxorig2} is convex and \eqref{eq:Wsum:noncvxorig1} is a DC constraint. Similarly to \eqref{eq:quad_over_lin}, we can use the first-order lower approximation for the right side of \eqref{eq:Wsum:noncvxorig1} as
\begin{equation}\label{eq:concave:1st}
 \tfrac{r_{b}^2}{z_b} \geq \tfrac{2r_{b}^{(n)}}{z_b^{(n)}}r_b - (\tfrac{r_b^{(n)}}{z_b^{(n)}})^2z_b \triangleq \varphi_b^{(n)}(r_b,z_b).
\end{equation}
With the linear approximations \eqref{eq:EEmax:quad_over_lin_app} and \eqref{eq:concave:1st}, we obtain an SCA-based iterative algorithm for solving \eqref{WsumEEmax} where the problem at iteration $n$ reads
\begin{subequations}
\label{eq:WsumEEmax:SCAproblem}
\begin{eqnarray}
\underset{\mathbf{t},\mathbf{r},\mathbf{z},\mathbf{w},\boldsymbol\gamma,\boldsymbol\beta}{\maxi} &  &   \sum_{b\in\mathcal{B}} \omega_bt_b\\
 \st &  &  t_b \leq \varphi_b^{(n)}(r_b,z_b), \forall b \in \mathcal{B}\label{eq:approx1}\\
&  &   \gamma_{k} \leq \Psi_k^{(n)}(\mathbf{w}_{k},\beta_{k}),\forall k \in \mathcal{K}\label{eq:approx2}\\
&  & \eqref{eq:WsumEE:PC},\eqref{eq:logalpha},\eqref{eq:Wsum:cvxorig2},\eqref{eq:interference:constraint}.
\end{eqnarray}
\end{subequations}
The proposed algorithm for the WsumEEmax problem is summarized in Algorithm \ref{algo:Alg2}. The same convergence discussions as in Algorithm \ref{algo:iterative} applies to Algorithm \ref{algo:Alg2} also. In Appendix \ref{app:SOCP}, we further show that the WsumEEmax problem can be approximated as an SOCP at each iteration. It is worth pointing out that the SCA-based method was also used in \cite{Tervo-15} to solve the energy efficiency maximization problem in a single-cell system. Herein, however, different transformations are used due to the rate dependent power consumption. The main differences are summarized in Appendix \ref{app:comparison}.

\begin{algorithm}[t]
\begin{algorithmic}[1] \caption{Proposed SCA-based beamformer design for the WsumEEmax problem.}
\label{algo:Alg2} \renewcommand{\algorithmicrequire}{\textbf{Initialization:}}
\REQUIRE Set $n=0$, and generate initial points $(\mathbf{w}^{(n)},\boldsymbol\beta^{(n)},\mathbf{z}^{(n)},\mathbf{r}^{(n)})$.
\REPEAT
\STATE Solve \eqref{eq:WsumEEmax:SCAproblem} with $(\mathbf{w}^{(n)},\boldsymbol\beta^{(n)},\mathbf{z}^{(n)},\mathbf{r}^{(n)})$
and denote optimal values as $(\mathbf{w}^*,\boldsymbol\beta^*,\mathbf{z}^*,\mathbf{r}^*)$.
\STATE Update $(\mathbf{w}^{(n)} = \mathbf{w}^* ,\boldsymbol\beta^{(n)} = \boldsymbol\beta^*,\mathbf{z}^{(n)}=\mathbf{z}^*,\mathbf{r}^{(n)}=\mathbf{r}^*)$.\label{algo:update}
\STATE $n:=n+1$.
\UNTIL desired accuracy
\renewcommand{\algorithmicrequire}{\textbf{Output:}} \REQUIRE $\mathbf{w}_k^*, \forall k \in \mathcal{K}$
\end{algorithmic}
\end{algorithm}

\subsection{Feasible Initial Points}
Finding a feasible initial point is an important issue for an SCA-based algorithm. For the NetEEmax problem, we can generate any beamformers $\mathbf{w}_{k}^{(0)}$ which satisfy the power constraints (which is easily done by normalization), then replace the equality (14d) with equality, i.e., calculate $\beta_k^{(0)}= N_0+\sum_{j\in\mathcal{K} \setminus \{k\}}|\mathbf{h}_{b_j,k}\herm\mathbf{w}_{j}^{(0)}|^2$. The resulting initial points $\mathbf{w}^{(0)},\boldsymbol\beta^{(0)}$ are feasible.
 The feasible initial values of $\mathbf{w}^{(0)},\boldsymbol\beta^{(0)}$ are also feasible to the WsumEEmax problem.
However, due to the additional approximation in (23b), we also need to find initial $\mathbf{r}^{(0)},\mathbf{z}^{(0)}$. To this end, we can first calculate $\gamma_k=\Gamma_k(\mathbf{w}^{(0)})$ according to (20c), then $r_b^{(0)}=\sqrt{\sum_{k\in\mathcal{K}_b}\log(1+\gamma_k)}$ according to (20d), and $z_b^{(0)}=g_b(\tilde{\mathbf{w}_b}^{(0)})+\delta(r_b^{(0)})$ which result in feasible initial points $\mathbf{w}^{(0)},\boldsymbol\beta^{(0)},\mathbf{r}^{(0)},\mathbf{z}^{(0)}$.

\section{Proposed Decentralized Solutions}
\label{sec:DecentralizedMethods}
The algorithms presented in the preceding section require
centralized processing. For the WsumEEmax problem, the rate function $\tilde{R}_b(\mathbf{w})$ in \eqref{eq:WsumEE:obj} couples all the cells due to the inter-cell interference (in the approximated problem, inter-cell interference appears in constraint \eqref{eq:interference:constraint}). In the NetEEmax problem, also the total power consumption in the objective function \eqref{eq:EE:obj} (the objective \eqref{eq:EEmax:reform4obj} in the approximated problem) couples all the cells. Due to the special structure of the WsumEEmax problem, we propose an alternative decentralized formulation which can be solved efficiently only relying on local channel state information and (scalar) backhaul information exchange. In particular, the proposed approach admits closed-form solutions and, thus, can be solved without invoking a generic external convex
solver. We will also discuss the challenge and also possibility to solve the
NetEEmax problem in a decentralized manner in the end of the section.

Let us start from problem \eqref{eq:WsumEEmax:reform3}. By adding 1 to both sides of \eqref{eq:SINRnonconvex} and using fact that when optimal MMSE receiver is used, it holds that $1+\Gamma_k(\mathbf{w})=\tfrac{1}{\epsilon_k(\mathbf{w},u_k)}$, where 
\begin{equation}
\begin{aligned}
\epsilon_k(\mathbf{w},u_k) = & |u_k|^2\bigl(\sum_{j\in\mathcal{K}}|\mathbf{h}_{b_j,k}\herm\mathbf{w}_j|^2+N_0\bigr) \\ & - 2\Re(u_k\mathbf{h}_{b_k,k}\herm\mathbf{w}_k) + 1
\end{aligned}
\end{equation}
is MSE and
\begin{equation}
\label{eq:receiverFirst}
u_k = \Bigl(\sum_{j\in\mathcal{K}}\mathbf{h}_{b_j,k}\herm\mathbf{w}_j\mathbf{w}_j\herm\mathbf{h}_{b_j,k} + N_0\Bigr)^{-1}\mathbf{h}_{b_k,k}\herm\mathbf{w}_k
\end{equation}
is the MMSE receiver of user $k$ \cite{Shi:WMMSE:2011}. Then, we can equivalently rewrite \eqref{eq:WsumEEmax:reform3} as
{\begin{subequations}
\label{eq:WsumEEmax:MSEform1}
\begin{align}
&\underset{\mathbf{t},\mathbf{r},\mathbf{z},\mathbf{w},\boldsymbol\gamma,\mathbf{u}}{\maxi} \quad   \sum_{b\in\mathcal{B}} \omega_bt_b\\
&\hspace{30pt}\st \hspace{10pt}   t_b \leq \frac{r_{b}^2}{z_b}, \forall b \in \mathcal{B}\label{eq:epi3:mse}\\
& \hspace{60pt} z_b \geq g_b(\tilde{\mathbf{w}}_{b}) + P_{\text{RD}}\delta(r_b), \forall b \in \mathcal{B}\label{eq:Wsum:cvxorig2:mse} \\
& \hspace{60pt}  \epsilon_k(\mathbf{w},u_k) \leq \frac{1}{1+\gamma_k}
, \forall k \in \mathcal{K} \label{eq:MSEnonconvex}\\
&  \hspace{60pt}\alpha\sum_{k\in\mathcal{K}_b}\log(1+\gamma_{k}) \geq r_{b}^2, \forall b \in \mathcal{B}\label{eq:logalpha:mse}\\
& \hspace{60pt} \sum_{k \in \mathcal{K}_b}||\mathbf{w}_{k}||_{2}^{2}\leq P_b,\;\forall b \in \mathcal{B}\label{eq:23PowConst}.
\end{align}
\end{subequations}}The above problem is still nonconvex even for fixed receivers $u_k$. However, if the receivers $u_k$ are fixed, all the other constraints are convex while \eqref{eq:epi3:mse} and \eqref{eq:MSEnonconvex} are DC constraints. The convex right hand side of \eqref{eq:epi3:mse} can be linearized as in \eqref{eq:concave:1st}. To deal with \eqref{eq:MSEnonconvex}, we linearize $\tfrac{1}{1+\gamma_k}$ around the point $\gamma_k^{(n)}$ as
\begin{equation}
\tfrac{1}{1+\gamma_k} \geq (1+\gamma_k^{(n)})^{-1} - \tfrac{1}{(1+\gamma_k^{(n)})^{2}}(\gamma_k-\gamma_k^{(n)})\label{eq:noncvx}.
\end{equation}
As a result, problem \eqref{eq:WsumEEmax:MSEform1} to find the beamformers for fixed receivers can be approximated as
\begin{subequations}
\label{eq:WsumEEmax:MSEformConvex}
\begin{align}
&\underset{\mathbf{t},\mathbf{r},\mathbf{z},\mathbf{w},\boldsymbol\gamma}{\maxi} \quad   \sum_{b\in\mathcal{B}} \omega_bt_b\label{eq:WsumEEmax:MSEformConvex:objective}\\
&\hspace{10pt}\st \hspace{10pt}  t_b \leq \tfrac{2r_{b}^{(n)}}{z_b^{(n)}}r_b - (\tfrac{r_b^{(n)}}{z_b^{(n)}})^2z_b, \forall b \in \mathcal{B}\label{MSEt}\\
& \hspace{40pt} z_b \geq g_b(\tilde{\mathbf{w}}_{b}) + P_{\text{RD}}\delta(r_b), \forall b \in \mathcal{B} \label{MSEz}\\
& \hspace{40pt} \epsilon_k(\mathbf{w},u_k) \leq \tfrac{1}{(1+\gamma_k^{(n)})} - \tfrac{\gamma_k-\gamma_k^{(n)}}{(1+\gamma_k^{(n)})^{2}}, \forall k \in \mathcal{K}\label{eq:noncvx}\\
&  \hspace{40pt}\alpha\sum_{k\in\mathcal{K}_b}\log(1+\gamma_{k}) \geq r_{b}^2, \forall b \in \mathcal{B}\label{eq:MSElogalpha}\\
& \hspace{40pt} \sum_{k \in \mathcal{K}_b}||\mathbf{w}_{k}||_{2}^{2}\leq P_b,\;\forall b \in \mathcal{B}\label{MSEpowerconst}.
\end{align}
\end{subequations}
Thus, for fixed receivers, the above convex problem can be run until convergence using the SCA approach. However, the monotonic convergence of the objective function is guaranteed even if we solve problem \eqref{eq:WsumEEmax:MSEformConvex} only once after receiver update, then update the linearization point (\eqref{MSEt} and \eqref{eq:noncvx}), followed again by receiver update and continue the procedure until convergence \cite{Venkatraman-16}. We refer to this algorithm as centralized method in the numerical results (e.g., Fig. \ref{fig: OTAiterations}). Problem \eqref{eq:WsumEEmax:MSEformConvex} still requires centralized processing, since all the cells are coupled due to interference terms in the left hand side of \eqref{eq:noncvx}. To enable decentralized processing, we resort to Karush-Kuhn-Tucker (KKT) conditions \cite{Boyd:ConvexOpt:2004} of \eqref{eq:WsumEEmax:MSEformConvex}. The Lagrangian of \eqref{eq:WsumEEmax:MSEformConvex} can be written as
\begin{equation} \label{EEmaxMSE3}
\begin{aligned}
& L(\mathbf{w},\mathbf{t},\mathbf{z},\boldsymbol\gamma,\mathbf{a},\mathbf{c},\mathbf{d},\mathbf{f},\mathbf{s}) =  \\ & \sum_{b\in\mathcal{B}} \omega_bt_b - \sum_{b\in\mathcal{B}}a_b(t_b-\tfrac{2r_b^{(n)}}{z_b^{(n)}}r_b + (\tfrac{r_b^{(n)}}{z_b^{(n)}})^2z_b)  \\
&    - \sum_{b\in\mathcal{B}}c_b(\tfrac{1}{\eta}\sum_{k\in\mathcal{K}_b}||\mathbf{w}_k||_2^2+P_{\text{RD}}\delta(r_b) + P_{\text{CP},b} - z_b)  \\ &   - \sum_{k\in\mathcal{K}}d_k(|u_k|^2(\sum_{j\in\mathcal{K}}|\mathbf{h}_{b_j,k}\herm\mathbf{w}_j|^2+N_0) \\ &  - 2\Re(u_k\mathbf{h}_{b_k,k}\herm\mathbf{w}_k) + 1 - (1+\gamma_k^{(n)})^{-1}  \\ &    + (1+\gamma_k^{(n)})^{-2}(\gamma_k-\gamma_k^{(n)}))  \\ &  - \sum_{b\in\mathcal{B}}f_b(r_b^2 - \alpha\sum_{k\in\mathcal{K}_b}\log(1+\gamma_k)) \\ &   - \sum_{b\in\mathcal{B}}s_b(\sum_{k\in\mathcal{K}_b}||\mathbf{w}_k||_2^2 -  P_b)
\end{aligned}
\end{equation}
where $\mathbf{w},\mathbf{t},\mathbf{z},\mathbf{r},\boldsymbol\gamma$ are primal variables and $\mathbf{a},\mathbf{c},\mathbf{d},\mathbf{f},\mathbf{s}$ are dual variables related to constraints \eqref{MSEt} - \eqref{MSEpowerconst}.
The KKT conditions of \eqref{eq:WsumEEmax:MSEformConvex} can be written as
\begin{subequations} \label{EEmaxMSE3}
\begin{align}{} & \frac{\partial L}{\partial w_k} = \tfrac{1}{\eta}c_b\mathbf{w}_k + \sum_{j\in\mathcal{K}}d_j|u_j|^2\mathbf{h}_{b_k,j}\mathbf{h}_{b_k,j}\herm\mathbf{w}_k\nonumber \\
&   - u_k\mathbf{h}_{b_k,k} d_k + s_b\mathbf{w}_k = 0, \forall k \in \mathcal{K} \label{wderiv}\\
&   \frac{\partial L}{\partial t_b} =\omega_b - a_b = 0 \label{tderiv}, \forall b \in \mathcal{B}\\
&   \frac{\partial L}{\partial r_b} =\tfrac{2r_b^{(n)}}{z_b^{(n)}}a_b - c_bP_{\text{RD}}\delta'(r_b) - 2f_br_b = 0, \forall b \in \mathcal{B} \label{rderiv}\\
&   \frac{\partial L}{\partial \gamma_k} =\gamma_k + 1 - \tfrac{f_b\alpha(1+\gamma_k^{(n)})^2}{d_k} = 0, \forall k \in \mathcal{K} \label{gammaderiv}\\
&   \frac{\partial L}{\partial z_b} =c_b - a_b(\tfrac{r_b^{(n)}}{z_b^{(n)}})^2 = 0, \forall b \in \mathcal{B}\label{zderiv}\\
&   \mathbf{c} \geq 0,\mathbf{d} \geq 0,\mathbf{f} \geq 0,\mathbf{s} \geq 0,\mathbf{a} \geq 0 \\
&   a_b(t_b-\tfrac{2r_b^{(n)}}{z_b^{(n)}}r_b + (\tfrac{r_b^{(n)}}{z_b^{(n)}})^2z_b) = 0, \forall b \in \mathcal{B} \\
&   c_b(\tfrac{1}{\eta}\sum_{k\in\mathcal{K}_b}||\mathbf{w}_k||_2^2+P_{\text{RD}}\delta(r_b) + P_{\text{CP},b} - z_b) = 0, \forall b \in \mathcal{B} \\
&   d_k(|u_k|^2(\sum_{j\in\mathcal{K}}|\mathbf{h}_{b_j,k}\herm\mathbf{w}_j|^2+N_0)  - 2\Re(u_k\mathbf{h}_{b_k,k}\herm\mathbf{w}_k) \nonumber \\ & + 1 - (1+\gamma_k^{(n)})^{-1}  + (1+\gamma_k^{(n)})^{-2}(\gamma_k-\gamma_k^{(n)})) = 0,\nonumber \\ & \forall k \in \mathcal{K} \\
&   f_b(r_b^2 - \alpha\sum_{k\in\mathcal{K}_b}\log(1+\gamma_k)) = 0, \forall b \in \mathcal{B} \\
&   s_b(\sum_{k\in\mathcal{K}_b}||\mathbf{w}_k||_2^2 -  P_b) = 0, \forall b \in \mathcal{B}\\
& \eqref{MSEt},\eqref{MSEz},\eqref{eq:noncvx},\eqref{eq:MSElogalpha},\eqref{MSEpowerconst}.
\end{align}
\end{subequations}
In \eqref{rderiv}, $\delta'(r_b)$ is derivative of $\delta(r_b)$. Equations \eqref{tderiv} and \eqref{zderiv} immediately imply that $a_b=\omega_b$ and $c_b = \omega_b(\tfrac{r_b^{(n)}}{z_b^{(n)}})^2$. The beamformers can be solved from \eqref{wderiv} as
\begin{eqnarray}{} \mathbf{w}_k= d_ku_k(\sum\limits_{j\in\mathcal{K}}d_j|u_j|^2\mathbf{h}_{b_k,j}\mathbf{h}_{b_k,j}\herm + s_b\mathbf{I} + \tfrac{1}{\eta}c_b\mathbf{I})^{-1}\mathbf{h}_{b_k,k}\label{wsolution}
\end{eqnarray}
where $d_k$ and $s_b$ are dual variables related to MSE constraints \eqref{eq:noncvx} and power constraints \eqref{MSEpowerconst}. The dual variables $s_b$ are chosen to satisfy the power constraints \eqref{MSEpowerconst} using the bisection algorithm \cite{Komulainen-13}. The MMSE receivers $u_k$ are solved as given in \eqref{eq:receiverFirst}.
Since \eqref{MSEt}-\eqref{eq:MSElogalpha} hold with equality at the optimum, we can write
\begin{subequations}
\begin{align}
& \gamma_k = -\epsilon_k(\mathbf{w},u_k)(1+\gamma_k^{(n)})^2 + (1+2\gamma_k^{(n)}) \\
&  r_b = \sqrt{\alpha\sum_{k\in\mathcal{K}_b}\log(1+\gamma_k)}\\
& z_b = g_b(\tilde{\mathbf{w}}_{b}) + P_{\text{RD}}\delta(r_b) \\
& t_b = \tfrac{2r_{b}^{(n)}}{z_b^{(n)}}r_b - (\tfrac{r_b^{(n)}}{z_b^{(n)}})^2z_b.
\end{align}
\end{subequations}
The dual variable $f_b$ can be computed from \eqref{rderiv} as
\begin{equation}
\label{eq:fstep}
f_b = \tfrac{2\tfrac{r_b^{(n)}}{z_b^{(n)}}\omega_b - \omega_b(\tfrac{r_b^{(n)}}{z_b^{(n)}})^2P_{\text{RD}}\delta'(r_b)}{2r_b}.
\end{equation}
Since the dual variables $f_b$ and $d_k$ depend on each other in \eqref{gammaderiv}, one has to be fixed to optimize for the other. In the proposed method, $d_k$ is fixed to
evaluate $f_b$ using \eqref{eq:fstep}. By solving $d_k$ from \eqref{gammaderiv}, the dual variable $d_k^{(i)}$ at iteration $i$ is a point in the line segment between $d_k^{(i-1)}$ and $\tfrac{f_b^{(i)}(1+\gamma_k^{(n)})^2}{\gamma_k^{(i)} + 1}$ determined by using a diminishing or a fixed step size $\rho^{(i)} \in [0,1]$, i.e.,
\begin{equation}
d_k^{(i)} = d_k^{(i-1)} + \rho^{(i)}(\tfrac{f_b^{(i)}\alpha(1+\gamma_k^{(n)})^2}{\gamma_k^{(i)} + 1} - d_k^{(i-1)}).\label{eq:subgradient_d}
\end{equation}
As can be seen, $d_k^{(i)} = \tfrac{f_b^{(i)}\alpha(1+\gamma_k^{(n)})^2}{\gamma_k^{(i)} + 1}$ is satisfied when $d_k$ converges.

To summarize, the updates in the iterative algorithm are:
\begin{subequations}\label{eq:closedformupdates}
\begin{align}{} & \mathbf{w}_k^{(i)}= d_k^{(i-1)}u_k^{(i-1)}(\sum_{j\in\mathcal{K}}d_j^{(i-1)}|u_j^{(i-1)}|^2\mathbf{h}_{b_k,j}\mathbf{h}_{b_k,j}\herm \nonumber \\ & + s_b\mathbf{I} + \tfrac{1}{\eta}c_b^{(i-1)}\mathbf{I})^{-1}\mathbf{h}_{b_k,k} \label{mse:wupdate}\\
& u_k^{(i)} = (\sum_{j\in\mathcal{K}}\mathbf{h}_{b_j,k}\herm\mathbf{w}_j^{(i)}(\mathbf{w}_j^{(i)})\herm\mathbf{h}_{b_j,k}+ N_0)^{-1}\mathbf{h}_{b_k,k}\herm\mathbf{w}_k^{(i)}\label{mse:uupdate}\\
& \gamma_k^{(i)} = -\epsilon_k(\mathbf{w}^{(i)},u_k^{(i)})(1+\gamma_k^{(i-1)})^2 + (1+2\gamma_k^{(i-1)})\label{mse:gammaupdate}\\
& r_b^{(i)} = \sqrt{\alpha\sum_{k\in\mathcal{K}_b}\log(1+\gamma_k^{(i)})} \label{mse:rupdate} \\
& z_b^{(i)} = \tfrac{1}{\eta}\sum_{k\in\mathcal{K}_b}||\mathbf{w}_k^{(i)}||_2^2 + P_{\text{CP},b} + P_{\text{RD}}\delta(r_b^{(i)}) \label{mse:zupdate}\\
& t_b^{(i)} = \tfrac{2{r}_b^{(i-1)}}{{z}_b^{(i-1)}}r_b^{(i)} - (\tfrac{{r}_b^{(i-1)}}{{z}_b^{(i-1)}})^2z_b^{(i)} \label{mse:tupdate}\\
& f_b^{(i)} = \tfrac{2\tfrac{{r}_b^{(i-1)}}{{z}_b^{(i-1)}}\omega_b - \omega_b(\tfrac{{r}_b^{(i-1)}}{{z}_b^{(i-1)}})^2P_{\text{RD}}\delta'(r_b^{(i)})}{2r_b^{(i)}}\label{mse:fupdate}\\
& d_k^{(i)} = d_k^{(i-1)} + \rho(\tfrac{f_b^{(i)}\alpha(1+\gamma_k^{(i-1)})^2}{\gamma_k^{(i)} + 1} - d_k^{(i-1)}) \label{mse:dupdate}\\
& c_b^{(i)} = \omega_b(\tfrac{{r}_b^{(i-1)}}{{z}_b^{(i-1)}})^2.\label{mse:cupdate}
\end{align}
\end{subequations}
Note that \eqref{mse:tupdate}, \eqref{mse:fupdate}, \eqref{mse:dupdate}, and \eqref{mse:cupdate} involve also $r_b^{(i-1)}$, $z_b^{(i-1)}$, and $\gamma_k^{(i-1)}$, i.e., these values are not updated until for the next iteration due to the fixed SCA step.
The beamformer structure in \eqref{mse:wupdate} resembles the one provided by the weighted MMSE method for weighted sum rate maximization (WSRmax) \cite{Shi:WMMSE:2011,Kaleva-16}. Here, however, the beamformer involves additional scaling factor $\tfrac{1}{\eta}c_b$ to reflect the EE objective.
Also, the stream specific scaling factor $d_k$ reflects the EE utility (not just the inverse of MSE as for WSRmax). The proposed decentralized algorithm is summarized in Algorithm \ref{algo:Decentralized}.\footnote{The proposed method can be used directly to the case with multiantenna receivers as well.}

\begin{algorithm}[t]
\begin{algorithmic}[1] \caption{Proposed decentralized beamformer design for the WsumEEmax problem.}
\label{algo:Decentralized} \renewcommand{\algorithmicrequire}{\textbf{Initialization:}}
\REQUIRE Set $n=0$, and generate initial points $(\boldsymbol\gamma^{(n)},\mathbf{z}^{(n)},\mathbf{r}^{(n)})$.
\REPEAT
\STATE BS $b, \forall b$: Update $\mathbf{w}_k^{(i)}$ using \eqref{mse:wupdate} and transmit precoded downlink pilots.
\STATE User $k, \forall k$: Using the effective channel information, calculate $u_k^{(i)}$ using \eqref{mse:uupdate}
\STATE User $k, \forall k$: Signal $u_k^{(i)}$ to BSs using precoded pilots
\STATE BS $b, \forall b$: Evaluate $\gamma_k^{(i)}, \forall k \in \mathcal{K}_b$, $r_b^{(i)}$, $f_b^{(i)}$, $z_b^{(i)}$, $t_b^{(i)}$, $c_b^{(i)}$ and $d_k^{(i)}, \forall k \in \mathcal{K}_b$
\STATE BS $b, \forall b$: Share $d_k^{(i)}, \forall k \in \mathcal{K}_b$ to other BSs via backhaul
\STATE $n:=n+1$.
\UNTIL convergence or $n >$ predefined maximum number of iterations.
\renewcommand{\algorithmicrequire}{\textbf{Output:}} \REQUIRE $\mathbf{w}_k^*, \forall k \in \mathcal{K}$
\end{algorithmic}
\end{algorithm}

\subsubsection*{Required Signaling}

The beamformer update in \eqref{mse:wupdate} requires information of the effective channels $u_j\mathbf{h}_{b_k,j}$ and dual variables $d_k$. The effective channels $u_j\mathbf{h}_{b_k,j}$ can be signaled to the BSs using precoded pilots, i.e., the UL pilots (sequences) are multiplied by the complex scalars (or vectors in multi-antenna case) \cite{Komulainen-13}. Due to the fact that the power consumption of BS $b$ depends on all the beamformers of BS $b$, $d_k$ can be only evaluated at the BS as such. In this case, $d_k$ should be shared via backhaul signaling to other BSs. This means that each BS needs to send $K_b$ scalar values per iteration to other BSs via backhaul. In a centralized method, a central controller requires the global channel information, i.e., all the channel vectors in the network. Each complex channel coefficient consists of two scalar coefficients. Let us assume an equal number of $L$ users per cell and $N$ antennas at each BS. Each BS needs to share $2(B-1)LN$ scalars to a central controller which performs all the processing, and it is $2(B-1)N$ times the decentralized case. It is worth observing that the signaling overhead of the centralized method scales both with the number of BS antennas $N$ and the number of users $L$, while only with $L$ for the decentralized method. Thus, if the number of antennas is large, the decentralized method requires less signaling even when the channels are static for longer time. However, the benefits of the decentralized method become more important in the time-varying channels because the BSs can easily acquire local channel information, and they only need to exchange the weights $d_k$ \cite{Tolli-11}. The centralized method would require sharing all the channel information every time when the channels change, which causes significant signaling overhead and delays.
 Another option would be to signal $f_b$ to the users of own cell, and then $d_k$ to BSs using uplink precoded pilots which could be done without any backhaul exchange. However, this would incur additional pilot resources. In practice, it can be more beneficial to perform multiple beamformer updates for fixed $u_k$'s to improve the convergence speed in terms of over-the-air iterations and reduce the signaling load. We study the convergence behaviour numerically in Fig. \ref{fig: OTAiterations}.

\subsubsection*{Notes on Convergence of Algorithm \ref{algo:Decentralized}}
For the centralized algorithm (see discussions after \eqref{eq:WsumEEmax:MSEformConvex}), the monotonic convergence of the objective function \eqref{eq:WsumEEmax:MSEformConvex:objective} can be guaranteed similarly to the queue deviation problem in \cite{Venkatraman-16}. Because we solve the KKT conditions of \eqref{eq:WsumEEmax:MSEformConvex} iteratively, the convergence of the distributed method could be guaranteed in a one special case. Specifically, we should first fix the receivers and linearization point ($\boldsymbol\gamma^{(n)}, \mathbf{z}^{(n)}, \mathbf{r}^{(n)}$), solve KKT conditions \eqref{EEmaxMSE3} (with fixed or diminishing step size $\rho$ in \eqref{eq:subgradient_d}) until convergence, update linearization point and receivers again, and continue the procedure by solving KKT conditions with fixed receivers and linearization point. In fact, this kind of method would be equivalent to the centralized algorithm. However, since we combine all the updates into a single iteration loop (i.e., update of linearization point, receivers, and the iterative line search for the dual variables $d_k$) to improve the convergence speed, the formal convergence of the objective cannot be guaranteed. Fortunately, we have experimentally observed the convergence of the objective function to the centralized solution if the step size $\rho$ is properly chosen. The choice of $\rho$ in \eqref{eq:subgradient_d} also affects the convergence speed of the algorithm. In the simulations, we set $\rho=0.15$. The convergence behaviour is numerically illustrated in Fig. \ref{fig: OTAiterations}. 
\begin{rem}\label{rem4}
Note that in order to enable decentralized implementation, BSs and users should alternately optimize the transmitters and receivers. If this is done sequentially, we need more pilot resources which reduces the available (time) resources for actual data transmission. Thus, there is a trade-off between the available over-the-air iterations and EE performance, depending on the coherence time of the channel (i.e., the size of the coherence block). In practice, we have to define the maximum number of over-the-air iterations which leaves sufficient amount of resources for data transmission \cite{Jayasinghe-15}.
\end{rem} 

\subsubsection*{Network EEmax problem}
The problem of network EEmax is more challenging to implement in a decentralized manner, due to the fact that both sum rate and sum power couple the entire network since they appear in a single fraction in \eqref{EEmax}. However, by allowing more backhaul signaling, it is possible to arrive at a slightly modified method of Algorithm \ref{algo:Decentralized}. Specifically, for the network EEmax, we replace \eqref{eq:WsumEEmax:MSEform1} as
\begin{subequations}
\label{eq:EEmax:MSEform1}
\begin{align}
&\underset{\mathbf{t},\mathbf{r},\mathbf{z},\mathbf{w},\boldsymbol\gamma,p,\mathbf{u}}{\maxi} \quad    \sum_{b\in\mathcal{B}}t_b\label{eq:EEmax:MSEform1obj}\\
&\hspace{30pt}\st \hspace{10pt} t_b \leq \frac{r_{b}^2}{p}, \forall b \in \mathcal{B}\label{eq:EEmax:Coupling}\\
& \hspace{30pt} \hspace{10pt}  p \geq \sum_{b\in\mathcal{B}}z_b\\
& \hspace{30pt} \hspace{10pt} \eqref{eq:Wsum:cvxorig2:mse}-\eqref{eq:23PowConst}
\end{align}
\end{subequations}
where we have added new variable $p$ for the ease of notation. Problem \eqref{eq:EEmax:MSEform1} is similar to the one in \eqref{eq:WsumEEmax:MSEform1} except now the sum power $p$ appears in constraint \eqref{eq:EEmax:Coupling} and, thus, links all the cells. By following the same steps to arrive at the updates in \eqref{eq:closedformupdates}, all the other updates remain the same except $a_b=1$, and $t_{b}^{(i)}, f_{b}^{(i)}$ and $c_b^{(i)}$ in \eqref{mse:tupdate}, \eqref{mse:fupdate} and \eqref{mse:cupdate} are replaced with
\begin{subequations}\label{eq:NetEEclosedformupdates}
\begin{align}{} & t_b^{(i)} = \tfrac{2{r}_b^{(i-1)}}{p^{(i-1)}}r_b^{(i)}-(\tfrac{{r}_b^{(i-1)}}{p^{(i-1)}})^2p^{(i)}\label{mseNetEE:tupdate}\\
& f_b^{(i)} = \tfrac{2\tfrac{{r}_b^{(i-1)}}{p^{(i-1)}} - c_b^{(i-1)}P_{\text{RD}}\delta'(r_b^{(i)})}{2r_b^{(i)}}\label{mseNetEE:fupdate}\\
& c_b^{(i)} = \sum_{b\in\mathcal{B}}(\tfrac{r_b^{(i-1)}}{p^{(i-1)}})^2,\label{mseNetEE:cupdate}
\end{align}
\end{subequations}
respectively. Furthermore, we need to add update for $p$ as
\begin{equation}
p^{(i)} = \sum_{b\in\mathcal{B}}z_b^{(i)}.\label{mseNetEE:pupdate}
\end{equation}
Equation \eqref{mseNetEE:cupdate} implies that $c_b^{(i)}$ is actually the same for all the BSs, so $c_b$ in the beamformer update \eqref{mse:wupdate} reflects the total network EE, instead of BS specific EE as in the WsumEEmax problem. We can see that due to the coupling in \eqref{mseNetEE:cupdate} and \eqref{mseNetEE:pupdate}, the BS specific scalars $(\tfrac{r_b^{(i-1)}}{p^{(i-1)}})^2$ and power consumption values $z_b$ need to be exchanged between the BSs. This is additional signaling overhead compared to the WsumEEmax problem. Although we have numerically observed the convergence of the method (as in the WsumEEmax problem), it may be difficult to implement in practice due to the extra signaling overhead. However, it is a numerically efficient method, which can be realized without invoking any optimization solver.

\section{Computational Complexity}\label{ComplexityComp}
Evaluating the computational complexity for Algorithms \ref{algo:iterative} and \ref{algo:Alg2} is challenging because they are based on solving optimization problems. Moreover, the worst-case complexity models (e.g., in \cite{Ben-Tal:2001:LMC:502969}) are in general rather conservative and do not give a realistic view. However, for Algorithm \ref{algo:Decentralized} and the closed-form solution for the NetEEmax presented in previous section, we can estimate the per-iteration complexity due to their closed-form structure. Thus, we use those as a baseline schemes for the complexity analysis.
The proposed algorithms basically result in a similar beamformer structure (see, e.g., Eq. \eqref{wsolution}) as the MMSE precoding in \cite{Bjornson-15}.
In fact, the expression in \eqref{wsolution} is a weighted MMSE beamformer, where the weights $d_k$ reflect the EE objective. The algorithm complexity is dominated by the beamformer expression, i.e., calculation of all the vector-vector multiplications and the matrix inversions. Thus, we can approximate the per-BS power consumption resulting from the beamformer optimization per iteration as
\begin{equation}
P_\text{LP,iter}^{\textrm{Alg. 3}} = \frac{W}{U}(\frac{N_b^3}{3L_{\text{BS}}} + \frac{3|\mathcal{K}|N_b^2 + 2N_b^2|\mathcal{K}_b| + \mathcal{O}(|\mathcal{K}|)}{L_{\text{BS}}}).\label{eq:Complexity}
\end{equation}
The above expression is different from the one used in \cite{Bjornson-15}. Specifically, the work of \cite{Bjornson-15} assumed massive MIMO setup with $|\mathcal{K}| < N_b$. In this paper, $|\mathcal{K}| > N_b$ throughout the simulations, and, thus, the zero-forcing beamformer does not exist. In the above expression, the power consumption scales cubicly with the number of antennas per BS in contrast to the number of users as in \cite{Bjornson-15}. Another difference is that a single-cell system was considered in \cite{Bjornson-15}, while in the multi-cell case considered herein, $|\mathcal{K}|$ in the expression $P_\text{LP,iter}$ is the total number of users in the network. $P_\text{LP,iter}$ assumes that the beamformer equation is solved by the standard Cholesky decomposition and forward/backward substitutions \cite{Boyd:Lectures:NumLinAlg}.
It is interesting to note that in fact the vector-vector multiplications $\mathbf{h}_k\herm\mathbf{h}_k$ (which cause the term $3|\mathcal{K}|N_b^2$ in \eqref{eq:Complexity}) dominate the complexity when $|\mathcal{K}|>N_b$, because it is the term scaling with the total number of users in the network.  All the other related computations are of linear complexity $\mathcal{O}(|\mathcal{K}|)$ and small compared to the other terms. It is worth observing that the complexities of the decentralized methods are significantly smaller than in Algorithms \ref{algo:iterative} and \ref{algo:Alg2}, which makes them tractable for energy-efficient processing.

\section{Pilot Allocation Strategy}\label{sec:PilotAllocation}
The pilot contamination in the network can be reduced by using more orthogonal pilot resources. However, this in turn decreases the available resources for data transmission. That is, there exists a trade-off between achieved energy efficiency and the number of pilot resources. In this section, we present a simple heuristic pilot allocation algorithm, which can achieve the energy efficiency trade-off explained above. A pilot allocation strategy has a significant impact on the pilot contamination, but a more detailed study of this topic is left for future work due to the space limitation.

If channel state information is perfectly known, the optimal pilot allocation could be found by exhaustive search, i.e., solving the energy efficiency optimization problem (network EEmax or WsumEEmax) for each possible pilot allocation combination. However, this is a combinatorial problem and intractable for a large network size. Moreover, in practice the optimization problem cannot be solved to find the best allocation because channel information is not known prior to pilot transmission. Thus, a pragmatic goal for the pilot contamination problem is to find a good pilot allocation strategy based on other prior information. In this paper, we propose a heuristic energy-efficient pilot allocation scheme which works as follows. If $\tau$ pilot resources are available (either in uplink or downlink), we assume that the users are divided into two different group sizes, $M_\text{max}=\ceil*{\tfrac{|\mathcal{K}|}{\tau}}$ and $M_\text{min}=M_\text{max}-1$, respectively. In each group of size $M_\text{min}$ and $M_\text{max}$, $M_\text{min}$ and $M_\text{max}$ users share one pilot resource, respectively. To minimize the effect of pilot contamination, the number of smaller size groups $X_\text{min}$ should be made as large as possible. Explicitly, we set $X_\text{max}=|\mathcal{K}|-(M_\text{max}-1)\tau$ to be the number of larger size groups and $X_\text{min}=(|\mathcal{K}|-X_\text{max}M_\text{max})$. For example, if $|\mathcal{K}|=21$ and $\tau=12$, then $M_\text{max}=2, M_\text{min}=1, X_\text{max}=9$, and $X_\text{min}=3$, i.e., 3 users get orthogonal pilot resource and the others are divided into pairs. The first step is to find a good strategy to allocate the users in the smaller groups which can be expected to have larger impact on the total energy efficiency. In order to get a good guess of the best users, we form all the possible user combinations of size $M_\text{min}$ and evaluate initial group specific energy efficiency metrics for each group based on known information (if $M_\text{min}=1$, then these are user-specific metrics). Towards this end, let us define the group specific energy efficiency metric for group $\mathcal{J}_l$ as
\begin{equation}
\label{eq:UserEEmetric}
\kappa_{l} = \frac{\sum_{k\in\mathcal{J}_l}\tilde{r}_k}{\tfrac{1}{\eta}\sum_{k\in\mathcal{J}_l}\tfrac{P_{b_k}}{|\mathcal{K}_{b_k}|}+|\mathcal{J}_l|\tfrac{\sum_{b\in\mathcal{B}}P_{\text{CP},b}}{|\mathcal{K}|}+P_{\text{RD}}\sum_{k\in\mathcal{J}_l}\tilde{r}_k^m}
\end{equation}
where $\tilde{r}_k \triangleq \alpha\log(1+\tfrac{\tfrac{P_{b_k}}{|\mathcal{K}_{b_k}|}\zeta_{b_k,k}}{\sum_{j\in\mathcal{J}_l \setminus \{k\}}\tfrac{P_{b_j}}{|\mathcal{K}_{b_j}|}\zeta_{b_j,k}+N_0})$, and $\zeta_{b_k,k}$ is the path gain (including shadowing) from BS $b_k$ to user $k$. Intuitively, we suggest to use $\tfrac{P_{b_k}}{|\mathcal{K}_{b_k}|}\zeta_{b_k,k}$ as a metric to indicate the desired channel power (assuming equal power allocation per user) and $\tfrac{P_{b_j}}{|\mathcal{K}_{b_j}|}\zeta_{b_j,k}$ the interfering channel power from the serving BS of user $j$ to user $k$. The second term in the denominator of \eqref{eq:UserEEmetric} represents the average circuit power for the user group $\mathcal{J}_l$. By the proposed metric, the first $X_\text{min}$ user groups are chosen based on having the largest $\kappa_{l}$'s. After the smaller size groups have been allocated, we again form all the possible user combinations of size $M_\text{max}$ from the remaining users which have not been allocated yet, and calculate the metrics \eqref{eq:UserEEmetric} for each group $\bar{\mathcal{J}}_l$. By following the idea of the first phase, the remaining pilot resources are allocated for the user groups having the largest $\kappa_{l}$ until all the users have been allocated. For the sake of completeness, the proposed energy-efficient pilot allocation strategy is presented in Algorithm \ref{algo:PilotAllocation}.

\begin{algorithm}[t]
\begin{algorithmic}[1] \caption{Proposed energy-efficient pilot allocation strategy.}
\label{algo:PilotAllocation}
\STATE Set $M_\text{max}=\ceil*{\tfrac{|\mathcal{K}|}{\tau}}$, $M_\text{min}=M_\text{max}-1$, $X_\text{max}=|\mathcal{K}|-(M_\text{max}-1)\tau$, $X_\text{min}=(|\mathcal{K}|-X_\text{max}M_\text{max})$, and form all the possible group combinations of size $M_\text{min}$.
\STATE For each group $\mathcal{J}_l$, evaluate energy efficiency metrics $\kappa_{l}$ according to \eqref{eq:UserEEmetric}.
\STATE Allocate the first $X_\text{min}$ user groups having the largest $\kappa_{l}$'s to the first $X_\text{min}$ pilot resources.
\STATE For the remaining users, form all possible combinations of user groups of size $M_\text{max}$, and evaluate the metrics \eqref{eq:UserEEmetric} for each group.
\STATE Allocate the remaining $X_\text{max}$ pilot resources to user groups having the largest $\kappa_{l}$'s.
\end{algorithmic}
\end{algorithm}

\section{Numerical Results}
\label{sec:NumericalResults}

We evaluate the performance for a quasistatic frequency flat Rayleigh fading channel model and consider 7-cell wrap-around model, where each user suffers interference from six neighboring base stations. We also assume a small-cell setup where the inter-BS distance $d_\text{B}$ is 120 m. The radius of each cell is assumed to be $\frac{d_\text{B}}{2}$, i.e., the cell edges are overlapping and the users are randomly dropped to the cell edges to enable fairness for the users. The path loss in dB is modeled as 35 + 30$\log(d)$ with distance $d$ in meters and the shadowing is modeled as log-normal distribution with a standard deviation of 8 dB \cite{LTETR12}. We set $|\mathcal{K}_b|=L, N_b = N$ for all $b$, i.e., $L$ is the number of users per cell and $N$ is the number of antennas at each BS, and the algorithms are stopped when the change over the last five iterations is smaller than $\xi=10^{-4}$. We set $\delta(y_b)=y_b^m$, where $y_b$ is the data rate of BS $b$ and $m\geq1$ is any rational number, and the other simulation parameters adopted from \cite{Earth:D2.3},\cite{Bjornson-15}, \cite{Wang-13}, \cite{Wang-15} are presented in Table \ref{tab:params}.

\begin{figure}[t]
\centering
\subfigure[Convergence illustration for three different initial points]{\label{fig:EEmaxConvAndCDFNatsSub1}\includegraphics[width=60mm]{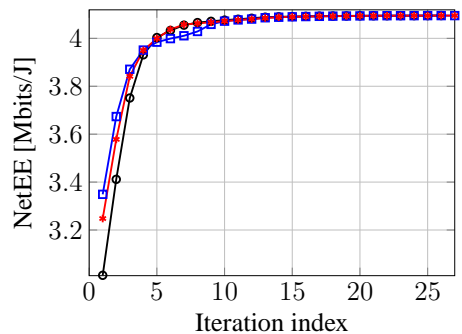}}
\subfigure[Average cumulative distribution function]{\label{fig:EEmaxConvAndCDFNatsSub2}\includegraphics[width=60mm]{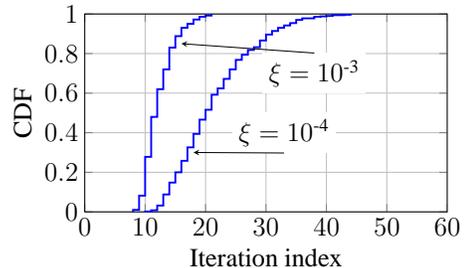}}
\caption{Convergence illustration of Algorithm \ref{algo:iterative} with $N=4$, $L=2$, $m=1.2$, $P_{\text{RD}}=2.4$ [W/(Gbits/s)$^m$].}
\label{fig:EEmaxConvAndCDFNats}
\end{figure}

\begin{figure}[t]
\centering
\subfigure[Convergence illustration for three different initial points]{\label{fig:WsumEEmaxConvAndCDFNatsSub1}\includegraphics[width=60mm]{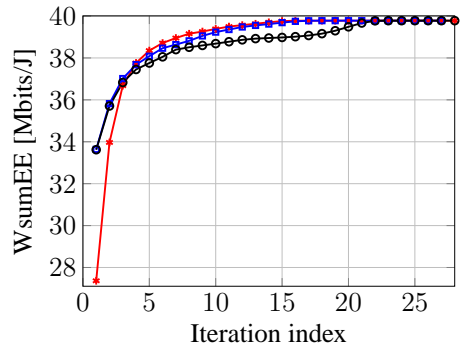}}
\subfigure[Average cumulative distribution function]{\label{fig:WsumEEmaxConvAndCDFNatsSub2}\includegraphics[width=60mm]{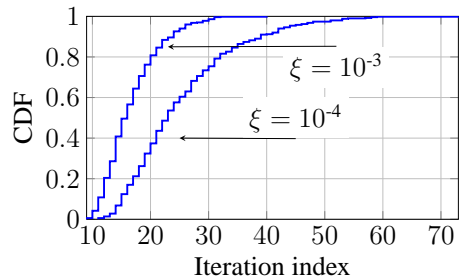}}
\caption{Convergence illustration of Algorithm \ref{algo:Alg2} with $N=4$, $L=2$, $m=1.2$, $P_{\text{RD}}=2.4$ [W/(Gbits/s)$^m$].}
\label{fig:WsumEEmaxConvAndCDFNats}
\end{figure}

\begin{figure}[t]
\centering
  \includegraphics[width=0.75\columnwidth]{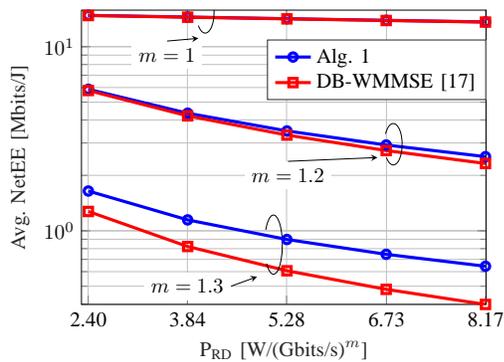}
  \caption{Energy efficiency comparison of the algorithms for different rate dependent power consumption models with $N=4$, $L=2$.}
  \label{fig: PrEffect}
\end{figure}

\begin{figure}[t]
\centering
\subfigure[$m=1$]{\label{fig:WsumPrEffectNatsSub1}\includegraphics[width=60mm]{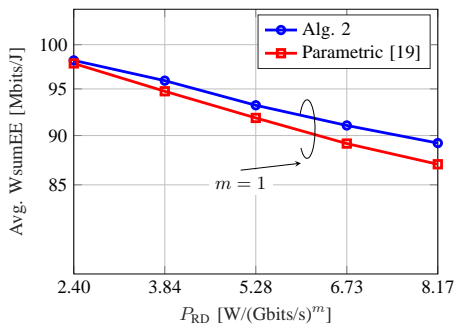}}
\subfigure[$m=1.2$ and $m=1.3$]{\label{fig:WsumPrEffectNatsSub2}\includegraphics[width=60mm]{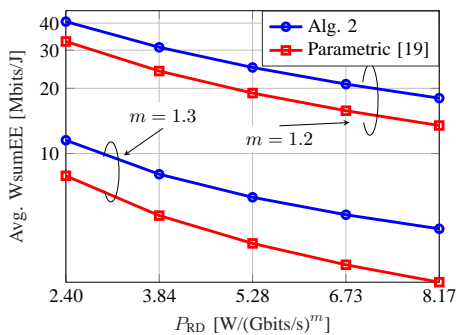}}
\caption{Weighted sum energy efficiency comparison of the algorithms for different rate dependent power consumption models with $N=4$, $L=2$.}
\label{fig: WsumPrEffectScales}
\end{figure}

\begin{figure}[t]
\centering
  \includegraphics[width=0.75\columnwidth]{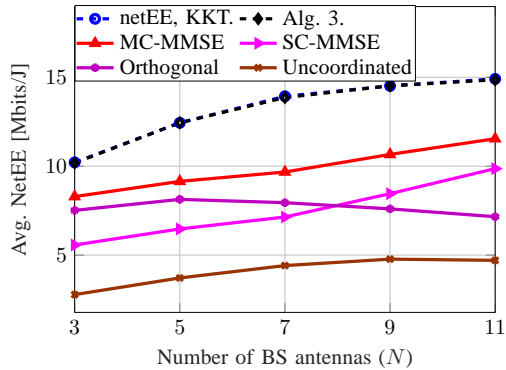}
  \caption{Average energy efficiency vs. the number of TX antennas $N$ with $L=3, P_{\text{RD}}=2.4$ [W/(Gbits/s)$^m$], $m=1$.}
  \label{fig: Antennas}
\end{figure}

\begin{figure}[t]
\centering
  \includegraphics[width=0.75\columnwidth]{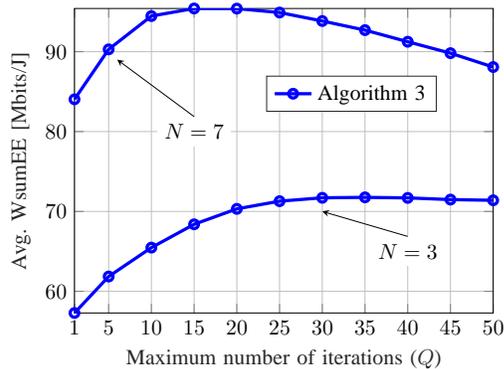}
  \caption{Average WsumEE vs. the maximum number of iterations $Q$ with $L=3, P_{\text{RD}}=2.4$ [W/(Gbits/s)$^m$], $m=1$.}
  \label{fig: EEvsQ}
\end{figure}

\begin{figure}[t]
\centering
  \includegraphics[width=0.75\columnwidth]{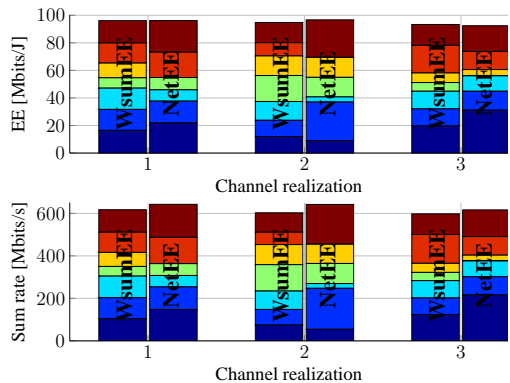}
  \caption{Comparison of individual energy efficiencies (upper figure) and sum rates (lower figure) of the cells achieved in network EEmax (right bar) and weighted sum EEmax (left bar) with $N=4$, $L=2$, $m=1$, $P_{\text{RD}}=2.4$ [W/(Gbits/s)$^m$]. The colors denote different base stations.}
  \label{fig: EERateBalance}
\end{figure}

\begin{figure}[t]
\centering
  \includegraphics[width=0.75\columnwidth]{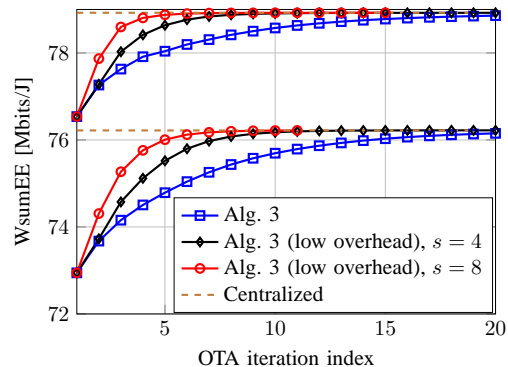}
  \caption{Convergence of Alg. \ref{algo:Decentralized} in terms of over-the-air iterations with $\rho=0.15, N=4, L=2, P_{\text{RD}}=2.4$ [W/(Gbits/s)$^m$], $m=1$.}
  \label{fig: OTAiterations}
\end{figure}

\begin{figure}[t]
\centering
  \includegraphics[width=0.75\columnwidth]{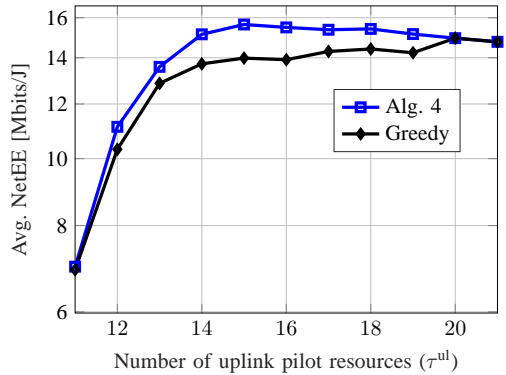}
  \caption{The effect of pilot contamination on the energy efficiency with $N=5, L=3, P_{\text{RD}}=2.4$ [W/(Gbits/s)$^m$], $m=1$.}
  \label{fig: PilotAllocation}
\end{figure}

In all the figures except \ref{fig: Antennas} and \ref{fig: EEvsQ}, we focus on algorithmic behavior, and, thus, use the same complexity model \eqref{eq:Complexity} for all the algorithms. In this regard, Figures \ref{fig: Antennas} and \ref{fig: EEvsQ} then demonstrate a more realistic view on the effect of complexity on the achieved EE of the proposed algorithms. In Figures \ref{fig:EEmaxConvAndCDFNats}-\ref{fig: OTAiterations}, we set $\tau^{\text{ul}}=\tau^{\text{dl}}=|\mathcal{K}|$, i.e., there is no pilot contamination, and in Figure \ref{fig: PilotAllocation}, the effect of pilot contamination is demonstrated.

\begin{table}[h]
\centering
\caption{Simulation parameters.}
\label{tab:params}
\begin{tabular}{ |c|c| }
\hline
Parameter & Value \\ \hline
 Inter-BS distance & 120 m\\
 Cell radius & 60 m \\
 $W$ & 20 MHz \\
 Path loss model & 35 + 30$\log(d \text{[m]})$ + $\mathcal{N}(0,8)$ dB \\
 $P_b$ & 27 dBm \\
 $U$ & 100 \\
 $\omega_b$ & 1 \\
 $N_0$ & -98 dBm \\
 $P_{\text{FIX}}$ & 3 Watt \\
 $P_{\text{BS}}$ & 0.4 Watt \\
 $P_{\text{SYN}}$ & 1 Watt \\
 $P_{\text{UE}}$ & 0.1 Watt \\
 $P_{\text{LP}}$ & Eq. \eqref{eq:PLP} \\
 $P_{\text{CE}}$ & 0.05 Watt \\
 $L_{\text{BS}}$ & 12.8 Gflops/W \\
 $Q$ & 20 \\
 $\eta$ & 0.2 \\ \hline
\end{tabular}
\end{table}


Figures \ref{fig:EEmaxConvAndCDFNats} and \ref{fig:WsumEEmaxConvAndCDFNats} plot the convergence and cumulative distribution functions (CDFs) of Algs. \ref{algo:iterative} and \ref{algo:Alg2}. We can see that both algorithms converge relatively fast in the considered setting and the obtained solution is insensitive to initial points. We have numerically observed that the proposed algorithms are stabilized (i.e., fairly close to the convergent point) after around 10 iterations. The convergence speed could be further increased by good initial points. We also show the cumulative distribution function of the number of iterations for different values of $\xi$ which also verifies a fast convergence speed.

Figure \ref{fig: PrEffect} illustrates the average energy efficiency as a function of $P_\text{RD}$ for different exponent values $m$. We compare the proposed algorithm with the existing method in \cite{He-13} (dubbed as "DB-WMMSE", referring to the combination of the Dinkelbach and WMMSE algorithm) where the rate dependent power consumption is not taken into account in the optimization problem. However, to have a fair comparison, after solving the energy efficiency problem with $P_\text{RD}=0$, the EE value plotted in Fig. \ref{fig: PrEffect} includes also the impact of rate dependent power. As analyzed mathematically in Remark \ref{Lemma1}, the rate dependent term does not affect the solution of the network EEmax when $m=1$. However, for a general model $m>1$, the proposed algorithm can provide up to 60\% gain in the considered setting, showing the importance of including the rate dependent power consumption in the optimization. Note that the gains of Alg. \ref{algo:iterative} depend on the setup. Larger gains can be achieved in the systems with low transmit power, where the rate dependent signal processing power consumption is a significant part of the total power consumption. Figure \ref{fig: PrEffect} also reveals that the power model has a huge impact on the energy efficiency which shows the importance of accurate power modeling.

Figure \ref{fig: WsumPrEffectScales} compares the WsumEE performance with different exponent values $m$ for the same simulation parameters as in Fig. \ref{fig: PrEffect}. Here, we similarly compare Alg. \ref{algo:Alg2} with the existing method in [19] (labelled by "Parametric") where the rate dependent power consumption is not taken into account in the optimization problem. We can see that for the WsumEE, Algorithm \ref{algo:Alg2} offers performance improvement compared to the traditional method even when linear rate dependent power consumption model is used (i.e., $m=1$).


Figure \ref{fig: Antennas} shows the average energy efficiency for the different numbers of transmit antennas $N$. Since the complexity evaluation is difficult for Algs. \ref{algo:iterative} and \ref{algo:Alg2} as discussed in Section \ref{ComplexityComp}, we plot the performance of the closed-form solutions presented in Section \ref{sec:DecentralizedMethods}. The closed-form solution for the NetEEmax presented in the end of Section \ref{sec:DecentralizedMethods} is labeled as \emph{`netEE, KKT'}. Note that in this figure, the actual network EE is also shown for Alg. \ref{algo:Decentralized}. Specifically, the beamformers are calculated with Alg. \ref{algo:Decentralized}, i.e., using the WsumEEmax as a design criterion, and the obtained beamformers are then used to calculate the objective of NetEEmax. 
We compare the proposed algorithms with various conventional beamforming methods such as the uncoordinated method, the orthogonal access method, and MMSE beamforming with multi-cell and single-cell processing. The uncoordinated and the orthogonal methods require no coordination, i.e., each BS tries to maximize its own EE. This means that their power consumption due to the complexity scales only with the number of users in each cell, i.e., $|\mathcal{K}|$ is replaced with $|\mathcal{K}_b|$ in \eqref{eq:Complexity}. In the uncoordinated method, all the BSs use all the bandwidth without any coordination, thereby causing severe inter-cell interference. In orthogonal access, the bandwidth is divided into 7 orthogonal sub-bands so that each BS occupies $W/7$ bandwidth, and, thus, the noise power level is also 7 times lower. In the MMSE precoding, we define $\mathbf{w}_k=\sqrt{p_k}\tilde{\mathbf{w}}_k$, where $p_k$ is the transmit power for user $k$ and $\tilde{\mathbf{w}}_k=(\mathbf{I}_N+\sum_{j\in\mathcal{K}}\frac{P_b}{LN_0}\mathbf{h}_{b_k,j}\mathbf{h}_{b_k,j}\herm)^{-1}\mathbf{h}_{b_k,k}/||(\mathbf{I}_N+\frac{P_b}{LN_0}\sum_{j\in\mathcal{K}}\mathbf{h}_{b_k,j}\mathbf{h}_{b_k,j}\herm)^{-1}\mathbf{h}_{b_k,k}||_2$ is the normalized MMSE beamforming vector. Then, the expressions $\tilde{\mathbf{w}}_k$ are plugged into problem (14), and the problem is iteratively solved by optimizing the power allocation to maximize the EE, while keeping $\tilde{\mathbf{w}}_k$ fixed. The \emph{multi-cell MMSE} optimizes the power allocation to maximize the network EE, while the \emph{single-cell MMSE} tries to maximize BS-specific energy efficiencies.
For the sake of fair comparison, it is assumed that the MMSE precoding method is a non-iterative method, i.e., $Q=1$ in the power consumption model \eqref{eq:Complexity}, while we fix $Q=20$ in the proposed algorithms. Thus, the MMSE consumes 1/20 fraction of the computational power of Alg. \ref{algo:Decentralized} and the `netEE, KKT' method. We can see that the energy efficiency increases with the number of antennas even though the power consumption due to the complexity (and also the number of RF chains) increases rapidly (see \eqref{eq:Complexity}). It is also observed that significant energy efficiency gains are achieved over all the other methods even though power consumption related to the computational complexity is higher. The large gains of the proposed methods over the MMSE precoding also reveal that optimizing the beamforming directions has a significant impact on the energy efficiency. The performance differences between the MMSE method and the proposed algorithms can be partly explained by looking, e.g., the MMSE beamformer structure and the optimized beamformer structure in \eqref{wsolution}. We can see that the MMSE beamformer gives equal weights to all the user channels, while the proposed beamformer structure maximizes the desired EE objective by adjusting the weighting towards each direction. More specifically, the relative importance of the cancelled/suppressed directions can be controlled by adjusting the user specific weights depending on the scenario. For example, in some scenarios it might be beneficial to assign (near) zero weights for a subset of users to maximize the EE objective. This would enable better interference controlling capability towards users with nonzero weights. This flexibility does not exist in the MMSE beamformer case, where the weights are assumed to be equal. In practice, there exists a trade-off between the number of iterations for the beamformer calculation and the achieved energy efficiency, because each iteration consumes some power but produces better beamformers toward the final ones. This trade-off is illustrated in the next experiment.

Fig. \ref{fig: EEvsQ} demonstrates the effect of algorithm complexity on the average energy efficiency of Algorithm \ref{algo:Decentralized}. More specifically, since each iteration of the algorithm consumes some amount power according to the power consumption model in \eqref{eq:Complexity}, we set maximum number of iterations $Q$ and stop the algorithm after $Q$ iterations which is shown in the x-axis. It is observed that the best number of iterations depends on the number of antennas. For example, the best WsumEE is achieved approximately in 15-20 iterations when $N=7$, while more iterations can be run for smaller number of antennas, because the complexity is lower. The performance decreases after a particular number of iterations due to the fact that the power consumption increase starts to dominate the achieved gain from the beamformer updates.

Figure \ref{fig: EERateBalance} compares individual base station energy efficiencies (upper figure) and sum rates (lower figure) achieved in the NetEEmax and WsumEEmax problems for three random channel realizations. As can be seen, when equal weights are used for all the BSs, the WsumEEmax clearly balances the energy efficiencies and rates between the cells with only small performance degradation in the network energy efficiency (also see Figure \ref{fig: Antennas} for further discussions). As far as fair resource allocation is concerned, the WsumEEmax design criterion proves to be a good choice.

Figure \ref{fig: OTAiterations} illustrates the convergence of Alg. \ref{algo:Decentralized} in terms of over-the-air (OTA) iterations for two different channel realizations. By OTA iteration we basically mean receiver update because it is a step which requires over-the-air downlink and uplink signaling. We also consider a variant of Alg. \ref{algo:Decentralized} (dubbed as "Alg. \ref{algo:Decentralized} (low overhead)" in Fig. \ref{fig: OTAiterations}) where after each receiver update, the beamformers are updated $s$ times at the BS side using only backhaul exchange. This means that steps 3-4 of Alg. \ref{algo:Decentralized} are performed less frequently, i.e., the pilot overhead due to precoded pilot transmission is reduced. As can be seen, all the variants of Alg. \ref{algo:Decentralized} converge to the centralized solution which is calculated by centralized alternating optimization between the receivers and beamformers (see the discussion after \eqref{eq:WsumEEmax:MSEformConvex}). We can see that the OTA signaling overhead can be reduced significantly with the variant method.


In Figure \ref{fig: PilotAllocation}, we illustrate the effect of pilot contamination on the network energy efficiency performance using a heuristic pilot allocation strategy provided in Section \ref{sec:PilotAllocation}, i.e., we use Algorithm \ref{algo:iterative} together with the pilot allocation algorithm. Let us recall that the pilot contamination in the network can be reduced by using more orthogonal pilot resources. However, this in turn decreases the available resources for data transmission. Thus, there exists an energy efficiency trade-off between the number of pilot resources and the effect of pilot contamination. Due to the limited resources, orthogonal allocation may not be possible in practice. Fortunately, it is unnecessary to use orthogonal resources for all the users because there may be some groups of users in which the users cause only small interference to each other possibly due to large spatial separation. Accordingly, these user groups could be allocated by the same resources to save resources for data transmission. This, however, requires advanced pilot allocation schemes. In Figure \ref{fig: PilotAllocation}, we focus on uplink pilot allocation, and simply keep the downlink pilot allocation orthogonal. The proposed Algorithm \ref{algo:PilotAllocation} is compared with the conventional "greedy" method, where the orthogonal users are first allocated based on maximum path gains (including shadowing), and then the remaining user pairs are allocated according to maximum sum path gain metrics, i.e., the metric $\kappa_{l}$ in steps 2-5 of Alg. \ref{algo:PilotAllocation} is replaced with $\sum_{k\in\mathcal{J}_k}\zeta_{b_k,k}$, and the user group having the largest $\kappa_{l}$ is allocated first. We can see that the proposed Algorithm \ref{algo:PilotAllocation} achieves better energy efficiency than the greedy method, and it is obtained with non-orthogonal allocation using $\tau^{\text{ul}}=$15 pilot resources. Note that when $\tau^{\text{ul}}=$15, 36\% of resources are wasted on DL and UL pilot signaling while the orthogonal allocation requires 42\% of the resources. This is the reason why reducing the UL pilot signaling can exceed the loss from less accurate beamforming. This interesting result verifies the existence of a trade-off between the number of pilot resources and the effect of pilot contamination with Algorithm \ref{algo:PilotAllocation}. When the available number of pilot resources is large, the greedy method gives only slightly worse performance than Alg. \ref{algo:PilotAllocation} because there are so many orthogonal users that it is sufficient to rely on path gains when allocating the resources. However, when the number of pilot resources becomes smaller, the performance of the greedy method is significantly inferior. Note that Fig. \ref{fig: PilotAllocation} is just a heuristic comparison of the pilot contamination effect, and a more detailed study would require its own body of work, which we would like to leave for future work. The existing state-of-the-art literature for this matter can be found, e.g., in \cite{Jose-12,Yin-13,Fernandes-13,You-15}.
\section{Conclusions}
\label{sec:Conclusions}
This paper has studied multi-cell energy-efficient coordinated beamforming with a rate dependent power consumption model. We have considered two different optimization criteria: network energy efficiency maximization and weighted sum energy efficiency maximization. The framework for the proposed solutions has been based
on the successive convex approximation principle. We have further proposed alternative formulations which enable decentralized closed-form implementations using only local channel state information and limited backhaul signaling. To reduce the effect of pilot contamination, a heuristic pilot allocation strategy has also been proposed in the paper. The numerical results have illustrated that the rate dependent power consumption has a significant impact on the energy efficiency and has to be taken into account when devising energy-efficient transmission strategies. The proposed methods have been shown to outperform various conventional beamformer designs. We have also demonstrated that the pilot contamination has a significant impact on the EE performance and showed that the proposed energy-efficient pilot allocation strategy can be used to achieve significant improvements when non-orthogonal pilot resources are used.

\appendices

\section{Required changes for alternative power consumption model}
\label{app:altPC}

In case the rate dependent power consumption of BS $b$ is modeled as $\sum_{k\in\mathcal{K}_b}P_{\text{RD}}\delta(r_k)$, where $\delta(r_k)$ is a function of individual user rate $r_k$, we replace $r_b$ with $r_k$ and apply this to the corresponding constraints so that the network EEmax problem \eqref{eq:EEmax:reformFinal} is replaced by
\begin{subequations}
\label{eq:EEmax:reformFinal_Altp}
\begin{align}
& \hspace{-4mm} \underset{\bar{\mathbf{w}},\bar{\boldsymbol\gamma},\bar{\boldsymbol\beta},\bar{\mathbf{r}},\phi}{\maxi} \quad \sum_{k\in\mathcal{K}}\bar{r}_k\\
\st \hspace{-2mm}  & \ \ \ \bar{r}_k - \phi\log(1+\tfrac{\bar{\gamma}_{k}}{\phi}) \leq 0, \forall k \in \mathcal{K}\\
& \hspace{4mm} \sum\limits_{k\in\mathcal{K}}(\tfrac{1}{\eta}\tfrac{||\bar{\mathbf{w}}_{k}||_2^2}{\phi} + P_{\text{RD}}\phi\delta(\tfrac{\bar{r}_k}{\phi})) + \phi \sum\limits_{b\in\mathcal{B}}P_{\text{CP},b} \leq 1 \\
& \hspace{4mm} \eqref{eq:CC_quad_over_lin}, \eqref{eq:IntConstForApp}, \eqref{eq:PowConstForApp}
\end{align}
\end{subequations}
where $\bar{\mathbf{r}}\triangleq \{r_k\}_{k\in\mathcal{K}}$. Following the same principle, the WsumEEmax problem in \eqref{eq:WsumEEmax:SCAproblem} is replaced by
\begin{subequations}
\label{eq:WsumEEmax:SCAproblem_Altp}
\begin{align}
& \hspace{-10mm} \underset{\mathbf{t},\mathbf{z},\mathbf{w},\boldsymbol\gamma,\boldsymbol\beta,\mathbf{r}}{\maxi} \quad   \sum_{b\in\mathcal{B}} \omega_bt_b\\
\st \hspace{5mm}  & \ t_b \leq \varphi_b^{(n)}(\{r_k\}_{k\in\mathcal{K}_b},z_b), \forall b \in \mathcal{B}\label{eq:approx1_Altp}\\
&\hspace{1mm} z_b \geq g_b(\tilde{\mathbf{w}}_{b}) + P_{\text{RD}}\sum_{k\in\mathcal{K}_b}\delta(r_k), \forall b \in \mathcal{B}\label{eq:Wsum:cvxorig2_Altp}\\
& \hspace{1mm} \log(1+\gamma_k) \geq r_k^2, \forall k \in \mathcal{K}\\
& \hspace{1mm} \eqref{eq:WsumEE:PC},\eqref{eq:interference:constraint}, \eqref{eq:approx2}
\end{align}
\end{subequations}
where $\varphi_b^{(n)}(\{r_k\}_{k\in\mathcal{K}_b},z_b)$ is now a first-order approximation of $\sum_{k\in\mathcal{K}_b}r_k^2/z_b$ instead of $r_b^2/{z_b}$.

\section{Iterative SOCP approximation of the WsumEEmax Problem}
\label{app:SOCP}
We note that \eqref{eq:Wsum:cvxorig2} is similar to \eqref{eq:EEmax:reformFinal:TotPower} and thus admits SOC representation if we consider a power model $\delta(y)=y^m$ as in Remark \ref{remPowerModel} (see Remark \ref{remPowerModel} for details). When this model is used, it is straightforward to show that all the constraints in \eqref{eq:WsumEEmax:SCAproblem} admit SOC form except the one in \eqref{eq:logalpha} which is an exponential cone. In our recent work \cite{Tervo-16}, we showed that an exponential cone can be approximated as a system of SOC constraints which results from the Taylor expansion (to several orders) of the exponential function. However, the number of slack variables introduced in such an approximation increases quickly with the number of users. Thus, herein we propose a novel way to approximate \eqref{eq:logalpha} as SOC constraint. In light of SCA principle, we need to find a concave lower bound for $\log(1+x)$. Let us denote $h(x)=-\log(1+x)$. Then we have
	\begin{multline}
	\left\Vert \nabla h(x_{1})-\nabla h(x_{2})\right\Vert _{2}=\left\lVert-\frac{1}{1+x_{1}}+\frac{1}{1+x_{2}}\right\rVert_{2}\\
	=\left\lVert\frac{x_{1}-x_{2}}{(1+x_{1})(1+x_{2})}\right\rVert_{2}\stackrel{(a)}{\leq}\left\lVert x_{1}-x_{2}\right\rVert_{2}\label{eq:prove:quadratic}
	\end{multline}
	for $x_1,x_2\geq0$. Note that  the inequality $(a)$ above is due to $(1+x_{1})(1+x_{2})>1$ for $x_{1},x_{2}>0$. That is to say, the gradient of $ h(x) $ is Lipschitz continuous with parameter $L=1$ on the domain $ x\geq 0 $. As a result, the function $ f(x)= \frac{L}{2}x^2-h(x) $ is convex, and thus we have \cite{Parikh-14}
\begin{equation}
f(y) \geq f(x) + \nabla f(x)^{T}(y-x)
\end{equation}	which is equivalent to
\begin{equation}
h(y)\leq h(x)+\nabla h(x)^{T}(y-x) +\frac{L}{2}(y-x)^2
\end{equation} for all $ x,y\geq 0 $. Substituting $ h(x)= -\log(1+x) $ into the above inequality leads to
\begin{equation}
\log(1+y)\geq \log(1+x)+\frac{1}{1+x}(y-x) -\frac{L}{2}(y-x)^2
\end{equation}
Now it is clear that we have the following inequality
\begin{equation} \label{approximation}
\begin{aligned}
\sum_{k\in \mathcal{K}_b} & \log(1+\gamma_k )  \geq  \sum_{k\in \mathcal{K}_b}\bigl(\log(1+\gamma^{(n)}_{k}) \\
& +\frac{1}{1+\gamma^{(n)}_{k}}(\gamma_k-\gamma^{(n)}_{k}) -\frac{L}{2}(\gamma_k-\gamma_k^{(n)})^2\bigr).
\end{aligned}
\end{equation}
Note the above inequality is tight and the first derivative of both sides with respect to $ \gamma_k $ are  equal when $ \gamma_k=\gamma_k^{(n)}, \forall k \in \mathcal{K}_b $, which satisfies the conditions of SCA framework. In summary, \eqref{eq:logalpha} can be approximated as
\begin{equation}\label{logSOCP}
 \sum\limits_{k\in \mathcal{K}_b}\alpha\bigl(\log(1+\gamma^{(n)}_{k})+\frac{\gamma_k-\gamma^{(n)}_{k}}{1+\gamma^{(n)}_{k}} -\frac{L}{2}(\gamma_k-\gamma_k^{(n)})^2\bigr) \geq r_{b}^{2}
\end{equation} which is an SOC constraint. Using the approximation \eqref{approximation}, we replace convex constraints \eqref{eq:logalpha} with \eqref{logSOCP} in \eqref{eq:WsumEEmax:SCAproblem}, and arrive at an SOCP in step 2 of Algorithm \ref{algo:Alg2}, which is much more efficiently solved than the generic convex formulation in \eqref{eq:WsumEEmax:SCAproblem}.

\section{Comparative Discussions to [6]}
\label{app:comparison}

In \cite{Tervo-15}, the successive convex approximation method was also used to solve the energy efficiency maximization problem in a single-cell system. In principle, one could extend the ideas from \cite{Tervo-15} to solve the problems considered in this paper. However, herein we use different transformations which have many advantages compared to the one in \cite{Tervo-15}, even when the rate dependent power would be ignored. Here we summarize the main differences.

Let us first consider the network EEmax problem. First, in \cite{Tervo-15}, the objective function is linearized in Eq. (22), followed by an introduction for new variables for the energy efficiency and the total power. In the proposed method, we only introduce new variables for the rate function in \eqref{eq:EEmax:reform1}, which implies that the proposed approach has fewer variables. Second, the linearization of the objective function in Eq. (22) of \cite{Tervo-15} also yields the fact that two different sets of approximated constraints are required in order to use the SCA method. In the proposed method, we use linear approximation for one set of constraints only, which can be expected to result in better performance in a general case. Third, in [6], the linear approximations are performed for the jointly concave geometric mean functions $\sqrt{ab}$. In the proposed method, we use linear approximations for the jointly convex quadratic-over-linear function $\frac{a^2}{b}$. In \cite{Tervo-15}, the right side approximation in (30b) can be written as
\begin{eqnarray}
\sqrt{(a-1)b} & \leq \sqrt{(a^{(n)}-1)b^{(n)}} + \frac{1}{2}(\frac{a^{(n)}-1}{b^{(n)}})(b-b^{(n)})\nonumber \\
& + \frac{1}{2}(\frac{b^{(n)}}{a^{(n)}-1})(a-a^{(n)})
\end{eqnarray}
In the above approximation, the denominator in $\frac{1}{2}(\frac{b^{(n)}}{a^{(n)}-1})$ can go to zero if the rate goes to zero, which can lead to a numerical problem (i.e., division by zero). In the proposed method, the denominator in \eqref{eq:EEmax:quad_over_lin_app} does not have this problem because $\beta_k$ (denoting the total interference-plus-noise) is always bounded below by the noise power. This is a particularly important point that needs to be accounted for in the considered problem, because we do impose no user-specific quality-of-service constraints.

In case of the WsumEEmax problem, since the objective function is a sum of fractional functions and the Charnes-Cooper transformation cannot be used, we linearize the objective function as done in \cite{Tervo-15}. However, we again approximate quadratic-over-linear functions instead of the geometric mean functions, which yields the fact that the constraints due to zero rates (as discussed above) cannot go to infinity in any case.
Also, in \cite{Tervo-15}, to approximate each convex problem as an SOCP, the exponential cone $x\geq e^{y}$ was approximated as a set of SOC constraints. However, this way introduces a lot of slack variables as we discuss in Appendix \ref{app:SOCP} above. In the proposed method presented in Appendix \ref{app:SOCP}, we propose a concave lower bound for $\log(1+x)$, which does not introduce any extra variables, meaning that the complexity is significantly reduced compared to the method in \cite{Tervo-15}.

\bibliographystyle{IEEEtran}

\end{document}